\documentclass[longnamefirst]{emulateapj}

\newcommand{\s}{s$^{-1}$}
\newcommand{\cm}{cm$^{-2}$}

\def\ltsima{$\; \buildrel < \over \sim \;$}
\def\simlt{\lower.5ex\hbox{\ltsima}}
\def\gtsima{$\; \buildrel > \over \sim \;$}
\def\simgt{\lower.5ex\hbox{\gtsima}}
\def\gsimeq
{\hbox{\raise0.5ex\hbox{$>\lower1.06ex\hbox{$\kern-1.07em{\sim}$}$}}}
\def\lsimeq
{\hbox{\raise0.5ex\hbox{$<\lower1.06ex\hbox{$\kern-1.07em{\sim}$}$}}}

\def\xmm{{\it XMM-Newton }}

\def\xmm{{\it XMM-Newton}}
\def\chandra{{\it Chandra}}
\def\suzaku{{\it Suzaku}}

\def\pg1404{PG~1404+226}
\def\sgras{Sgr~A$^{*}$}

\def\dchisq{\hbox{$\Delta\chi^2$}}

\def\mnras{MNRAS}
\def\apj{ApJ}
\def\aj{AJ}
\def\apjl{ApJL}
\def\apjs{ApJS}
\def\aap{A\&A}
\def\araa{ARA\&A}
\def\nat{Nature}
\def\pasj{PASJ}

\shorttitle{Discovery of superluminal Fe K echo}
\shortauthors{G.\ Ponti et al.}

\begin{document}
\title{Discovery of a superluminal Fe K echo at the Galactic Center: \\
 The glorious past of Sgr A* preserved by molecular clouds}

\author{G. Ponti\altaffilmark{1,2}, R. Terrier\altaffilmark{1}, A. Goldwurm\altaffilmark{1,3}, 
G. Belanger\altaffilmark{4} and G. Trap\altaffilmark{1,3}} 
\altaffiltext{1}{APC Universit\'e Paris 7 Denis Diderot, 75205 Paris Cedex 13, France}
\altaffiltext{2}{School of Physics and Astronomy, University of Southampton, Highfield, 
   Southampton, SO17 1BJ, UK}
\altaffiltext{3}{Service dÕAstrophysique/IRFU/DSM, CEA Saclay, Bat. 709, 91191 
   Gif-sur-Yvette Cedex, France}
\altaffiltext{4}{ESA/ESAC, PO Box 78, 28691 Villanueva de la Ca\~nada, Spain} 
 \email{ponti@iasfbo.inaf.it}

\begin{abstract}
We present the result of a study of the X-ray emission from the Galactic Centre (GC) 
Molecular Clouds (MC) within 15 arcmin from Sgr A*. We use \xmm\ data 
(about 1.2 Ms of observation time) spanning about 8 years.
The MC spectra show all the features characteristic of reflection: i) intense Fe K$\alpha$, 
with EW of about 0.7-1 keV, and the associated K$\beta$ line; ii) flat power law continuum 
and iii) a significant Fe K edge ($\tau\sim0.1-0.3$). The diffuse low ionisation Fe K emission 
follows the MC distribution, nevertheless not all MC are Fe K emitters. The long 
baseline monitoring allows the characterisation of the temporal evolution of the 
MC emission. A complex pattern of variations is shown by the different MC, with some 
having constant Fe K emission, some increasing and some decreasing. In particular,
we observe an apparent super-luminal motion of a light front illuminating 
a Molecular nebula. This might be due to a source outside the MC (such as Sgr A* or 
a bright and long outburst of a X-ray binary), while it cannot be due to low energy 
cosmic rays or a source located inside the cloud. 
We also observe a decrease of the X-ray emission from G0.11-0.11, behaviour similar 
to the one of Sgr B2. The line intensities, clouds dimensions, columns densities and 
positions with respect to Sgr A*, are consistent with being produced by the same 
Sgr A* flare.
The required high luminosity (about 1.5$\times$10$^{39}$ erg s$^{-1}$) can hardly be 
produced by a binary system, while it is in agreement with a flare of Sgr A* fading 
about 100 years ago. The low intensity of the Fe K emission coming from the 50 and 
the 20 km \s\ MC places an upper limit of 10$^{36}$ erg \s\ to the mean 
luminosity of Sgr A* in the last 60-90 years. The Fe K emission and variations from 
these MC might have been produced by a single flare of Sgr A*.

\end{abstract}


\section{Introduction}

Sgr A*, the supermassive black hole (BH) at the center of the Milky Way, 
now radiates at a rate 
about 8 orders of magnitude lower than the Eddington luminosity for its
estimated mass of M$_{BH}\sim$4~$\times10^6$ M$_{\odot}$ 
(Sch\"odel et al. 2002; Eisenhauer et al. 2003; Ghez et al. 2003; 2005; 
Gillesen et al. 2009). 
Such a low luminosity is difficult to reconcile with the dense environment 
that is present in the center of the Galaxy and has motivated the development 
of several radiatively inefficient accretion/ejection models (Melia \& Falcke 2001).
Although Sgr A* is known to display flares in X-rays (Baganoff et al. 2001; 
Goldwurm et al. 2003) and near-infrared (Genzel et al. 2003; Ghez et al. 2004), 
during which the X-ray intensity increases by factors up to 160 (Porquet et al. 2003) 
from the quiescent value, the bolometric luminosity still remains extremely low 
during these events compared to the Eddington one or even to the accretion 
power expected from the capture of stellar wind material from the nearby stars.
On the other hand, one may wonder whether Sgr A* has always been so 
underluminous or if it experienced, in the past, long periods of high energy activity,
that would make the massive black hole of our Galaxy more similar, than appears 
today, to typical low-luminosity Active Galactic Nuclei.

Indication of Sgr A* past activity can be sought in the interstellar medium 
surrounding the black hole. Sunyaev et al. (1993) were the first to interpret 
the X-ray emission, seen with GRANAT to roughly follow the distribution of the 
Molecular Clouds (MC) of the region, as scattering by the molecular material 
of emission from a past outburst of Sgr A* and predicted, at that time, a 
correlation of the X-ray fluorescent line of neutral iron with the MC.
Koyama et al. (1996) with ASCA and Murakami et al. (2001b) with Chandra
did in fact find such a correlation, particularly evident with the most massive 
MC complex of the region, Sgr B2, and proposed, using parameters derived 
from this cloud, that Sgr A* underwent, about 300 years ago, a major outburst 
of X-ray emission, with a luminosity of the order of few 10$^{39}$ ergs s$^{-1}$. 

The fluorescence line at 6.4 keV (K$\alpha$) is produced by the extraction of an
electron from the inner shell (K) of neutral or low-ionized iron atoms and the 
following electron transition from the second shell (L). 
Such line (actually a close doublet) is generally associated with another line (K$\beta$) 
due to the transition from the upper (M) level. 
Collisionally-ionized iron atoms in a hot plasma preferentially produce lines
in the 6.5-6.9 keV range, associated with a plasma continuum spectrum. 
Thus, the origin of the 6.4 keV line is most probably associated with either a 
large irradiation by photons having energies higher than 7.1 keV or by energetic 
particles, most probably electrons. 

Diffuse X-ray (2-10 keV) emission in the galactic center region is complex and 
still under intense investigation (Park et al. 2004; Goldwurm 2008; Koyama et al. 2009) 
but it certainly consists of at least the following components: a uniformly 
distributed soft emission well described by a low temperature ($\approx$1 keV) 
plasma, a little less uniform but centrally peaked 6.7 keV line associated with 
continuum emission described by a hot (kT $\approx 7$ keV) plasma model, 
and a clumpy 6.4 keV iron line component well correlated with molecular material.
The soft component can be fully explained by the supernova (SN) activity of 
the region, while the origin of the other components is more uncertain.
The interpretation of hot plasma emission for the 6.7 keV line and associated 
hard component is problematic because such plasma cannot be confined in 
the region and its regeneration would require a too large amount of energy
(but see Belmont \& Tagger (2006) for a heating mechanism of a helium 
dominated hot plasma at the galactic center).
An alternative is that the hot component may contain an important contribution 
from faint sources. This interpretation is supported by the similarity in the 
X-ray and near-infrared surface brightness distribution (Revnivtsev et al. 2006b) 
and by the large fraction of weak point sources in the population that the 
Chandra deep survey of the galactic center unveiled (Muno et al. 2004, 
Revnivtsev et al. 2007). 
For the 6.4 keV component, two models of emission are competing:
the reflection model quoted above (Sunyaev \& Churazov 1998), and one that 
attributes the 6.4 keV emission to low energy particles, most probably 
electrons (Valinia et al. 2000; Yusef-Zadeh et al 2002; 2007).
In the frame of the latter model, the impact by low energy protons (Dogiel et al. 2009) or 
by fast moving SN ejecta (Bykov 2003) with neutral material have also 
been considered.

This emission line has been detected in other, but not all, molecular clouds 
of the central region (Murakami et al. 2001a; Yusef-Zadeh et al. 2002; Nakajima
et al. 2009), for which a reflection from single Sgr A* event interpretation is 
more problematic and for some of which has also been found a correlation 
with non-thermal radio filaments, indicating local particle acceleration 
(Yusef-Zadeh et al. 2007).

On the other hand, the detection of hard X-ray emission up to 100 keV from Sgr B2 
obtained with the Integral observatory by Revnivtsev et al. (2004) supports 
the reflection nebula interpretation. 
These authors also demonstrated that the emission line intensity was constant 
until about 2000 and therefore that the original outburst must have lasted at
least 10 years. 
However, the most convincing evidence that supports the photon-ionisation model 
from an external source, comes now from the recent detections of variability of the line 
and continuum emission, a signature predicted and modelled in detail by Sunyaev 
\& Churazov (1998). Up to now two different claims of variability detection have 
been published. Muno et al. (2007) observed variation in the continuum (not in the 
line emission) flux and morphology from two 6.4 keV nebulae 
at about 6 arcmin from Sgr A*. Then using data from several satellites (ASCA, 
\chandra, \xmm\ and Suzaku), Koyama et al. (2008) and Inui et al. (2009) showed 
that the Sgr B2 6.4 keV line emission is changing in a way that it would be 
produced by a wave front passing through the different components of the 
Sgr B2 complex.

Maybe the more compelling evidence, up to now, is the discovery 
of time evolution of the hard X-ray emission from Sgr B2 observed 
over 7 years by the same instrument on the Integral satellite (Terrier et al. 2010).
Indeed the evolution of the hard X-ray emission of Sgr B2 is best explained by 
an X-ray reflection nebula scenario in which the fading of the reflection component 
(for the first time measured in the Compton hump region) is due to the propagation 
of the decay part of the outburst and rules out competing models based 
on irradiation by low energy cosmic ray electrons.

However, Sgr B2 is quite a special object and one may wonder whether the reflection
nebula model works only for this cloud or if it holds for the other 6.4 keV features of 
the region. If the origin of the scattered emission is external to the cloud, signatures 
should be seen elsewhere. Sgr B2 is indeed not the only MC close to Sgr A*. 
The supermassive BH sits on the middle of the Central Molecular Zone (CMZ; 
Morris \& Serabyn 1996), a condensation of MCs right in the center of the Galaxy.
The detailed study of this region can therefore validate the reflection model
and even narrate the past history of Sgr A* emission (Sunyaev et al. 1993; 
1998; Cramphorn et al. 2002). Moreover, the light front due to a major flare may be 
used as a tool to scan the distribution of the MC material in the CMZ.

The main focus of this work is the study of the X-ray emission from the molecular 
clouds located around Sgr A* using the 8-year Sgr A* monitoring program 
carried out with the \xmm\ satellite. 

In section 2 the different observations and data 
reduction are presented. Section 3 shows the GC images in the 
Fe K$\alpha$ band and the analysis of the CS maps, with the 
aim to correlate the Fe K emission with the MC disposition, so that 
we can infer the location/distribution and column density of the 
different MC within the \xmm\ field of view. Section 4 presents 
the mean spectra from the different regions selected through 
the CS maps. Section 5 shows the time evolution of the Fe K 
emission from the MC. Section 6 shows the discovery of a 
super-luminal echo in a X-ray reflecting nebula. 
In Section 7 the results are discussed, in particular in subsection 7.2 
the possibility that all MC are illuminated by a single flare from Sgr A* 
is discussed. The conclusions are summarised in Section 8.

\section{Observation and data reduction}

Sgr A* and the molecular clouds in its immediate proximity, i.e. within a 
circle of 15 arcmin radius, have been observed up to now 13 times by 
\xmm, spanning a period of 8 years.
Table \ref{Obs} shows the observation dates, exposures and IDs of the 
different pointings, all directed towards \sgras.
Photon list files were created from the original Observation Data Files
(ODF) using the \xmm\ Science Analysis Software (SAS)
version 7.1.0. The EPIC pn and both MOS cameras were operated 
in full--frame and large--window mode, respectively. 
The loss of CCD6 in the EPIC-MOS1 camera after 9 March 2005, 
(most probably due to a micro-meteoroid; Abbey et al. 2006), 
prevents us from extracting MC spectra from the corresponding 
region during the 2007, 2008 and 2009 observations.

Most of the observations were affected by strong particle background 
flares. In order to reduce such contamination, 
we visually inspected all the EPIC-pn light curves and cut the 
periods during which the pn full field of view 10-12 keV count rate is 
higher than about 2 counts/s. This severely reduced the exposure 
time (see Tab. \ref{Obs}), but it provides cleaned data sets, where the impact 
of the particle background is negligible. We then select the same 
good time intervals for the EPIC-MOS camera.
Both for pn and MOS cameras, single and double events have been selected.
For spectra we selected {\sc flag} zero events only, while the expression 
{\sc (flag \& 0xfb002c) == 0} has been used to extract photons to produce 
images.

For the spectra, source plus background photons were extracted from the 
regions shown in Fig. \ref{FeK_EW}. The background spectrum
was extracted from regions south of the Galactic plane where the impact 
of point sources is minimal (see right panel of Fig. \ref{FeK}).
With the SAS commands {\footnotesize ARFGEN} and {\footnotesize RMFGEN} 
ancillary and response files were created.
The source and background spectra and response files are added 
with the {\sc ftool mathpha}. 
In the subsequent spectral analysis, source spectra were grouped such that
each bin contained at least 30 counts and errors are purely statistical and 
quoted at the 90 per cent confidence level (\dchisq=2.7 for one interesting
parameter), if not otherwise specified. Spectral fitting was performed 
using XSPEC 12.4.0.

All the images have been first corrected for out of time events (produced with 
the SAS command and then subtracted using the {\sc ftools} command {\sc farith}). 
These images have been then merged with the {\sc sas} command 
{\sc emosaic}. The exposure maps were calculated with the command 
{\sc eexpmap} and then used to normalise the images. 
With {\sc ds9} a Gaussian smoothing, with a kernel radius of 10 pixels, has been 
applied to the resulting images that have 1 arcsec pixels.

\begin{table}
\small
\begin{center}
\begin{tabular}{ccc ccc c}
\hline 
\hline 
N  & Gr & Date            & Time & Expo & Obs. ID        & Filter\\
     &      &                    &   (ks) &  (ks)  & \\
1   & 1  & 2001-09-04 &   27.7 & 18.3 & 0112972101 & M \\
2   & 1  & 2002-10-03 &   17.3 &   8.6 & 0111350301 & T \\
3   & 2  & 2004-03-28 & 133.0 & 17.3 & 0202670501 & M \\
4   & 2  & 2004-03-30 & 134.4 & 35.4 & 0202670601 & M \\
5   & 3  & 2004-08-31 & 135.2 & 84.3 & 0202670701 & M \\
6   & 3  & 2004-09-02 & 135.2 & 94.6 & 0202670801 & M \\
7   & 4  & 2007-03-30 &   35.2 & 57.2 & 0402430701 & M \\
8   & 4  & 2007-04-01 & 105.4 & 39.4 & 0402430301 & M \\
9   & 4  & 2007-04-03 & 105.8 & 25.3 & 0402430401 & M \\
10 & 5  & 2008-03-23 & 105.7 & 72.4 & 0505670101 & M \\
11 & 6  & 2009-04-01 &   39.9 & 28.8 & 0554750401 & M \\
12 & 6  & 2009-04-03 &   44.3 & 35.5 & 0554750501 & M \\
13 & 6  & 2009-04-05 &   39.1 & 27.8 & 0554750601 & M \\
\hline 
\hline 
\end{tabular}
\end{center}
\caption{Progressive observation number, progressive number of the summed 
observations, observation date, total observation time, cleaned exposure (EPIC-pn), 
observation ID and filter of the \xmm\ observations pointed towards Sgr A*.}
\label{Obs}
\end{table} 

\section{Spatial distribution of the Fe K$\alpha$ diffuse emission}

\begin{figure*}
\includegraphics[width=0.46\textwidth,height=0.495\textwidth,angle=0]{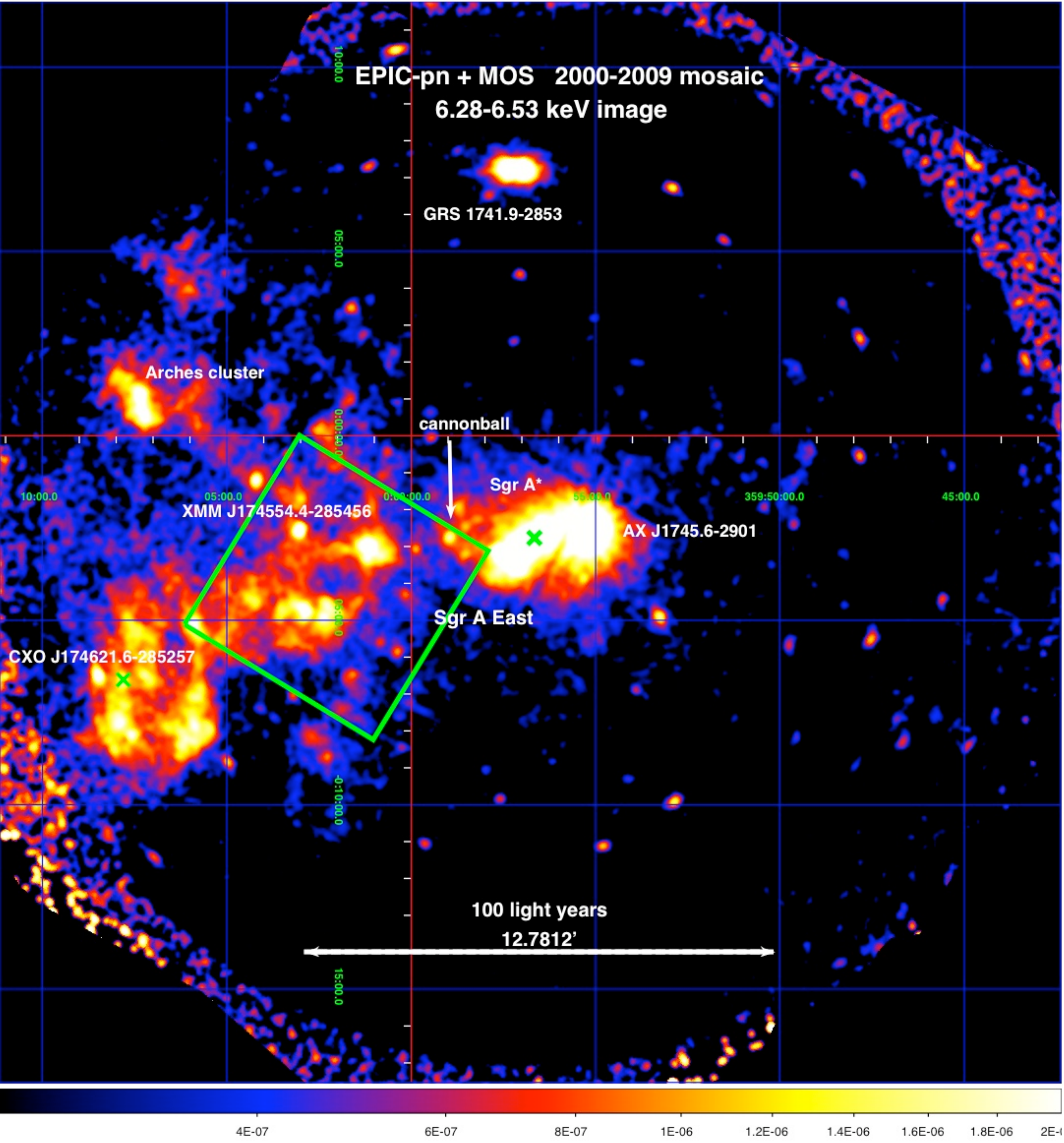}
\includegraphics[width=0.46\textwidth,height=0.495\textwidth,angle=0]{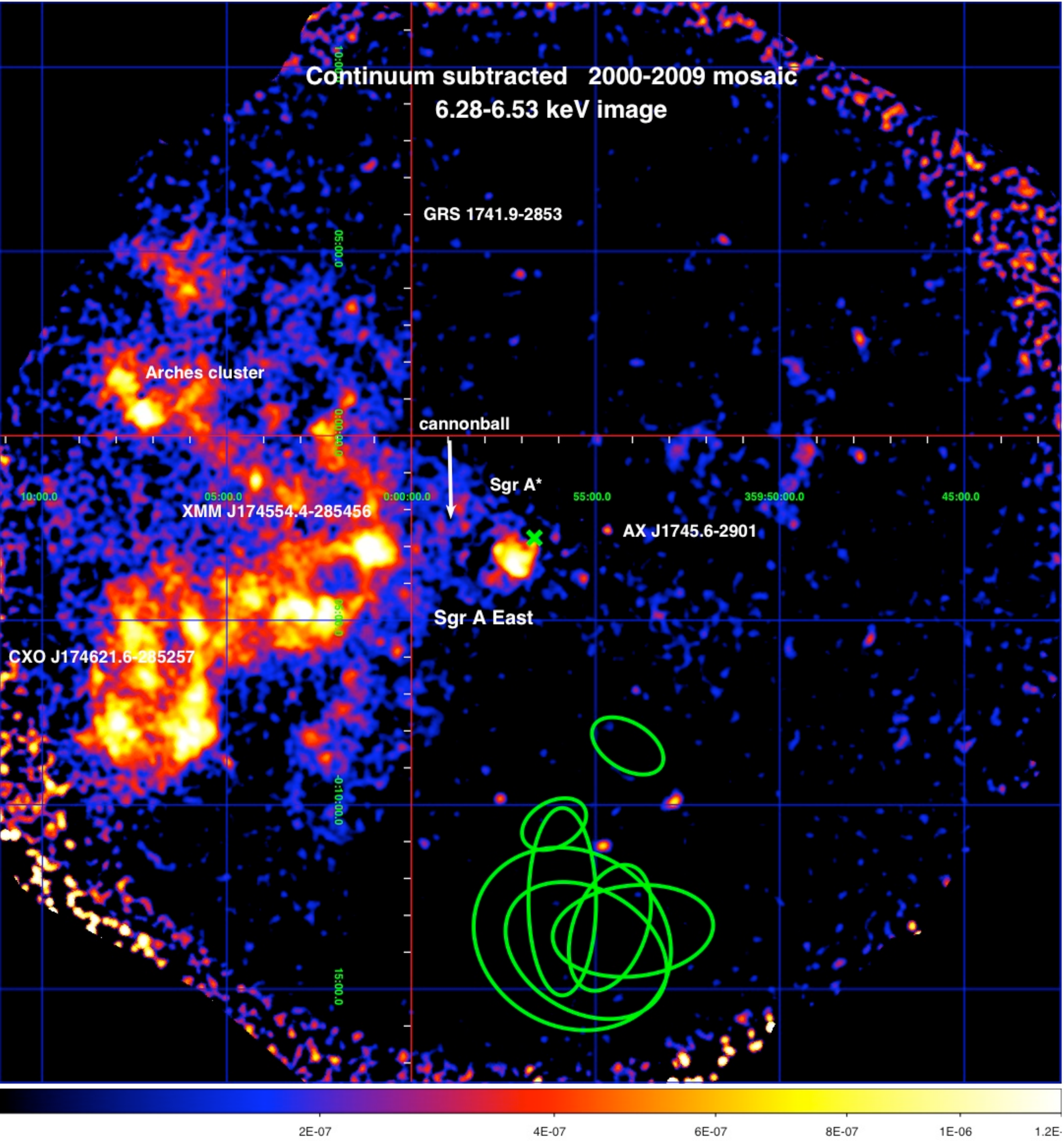}
\caption{{\it (Left panel)} Mosaic EPIC-pn + MOS images of all the \xmm\ 
observation between 2000 and 2009. The events are selected between 
6.28 and 6.53 keV, the energy band that contains the neutral Fe K emission line.
The stronger features are due to transient point sources with strong continuum and 
weak Fe K$\alpha$ lines. The green square shows the region studied by Muno 
et al. (2007) and shown in their Fig. 2.
{\it (Right panel)} Continuum subtracted Fe K image assuming an absorbed 
power law continuum shape with spectral index of $\Gamma=2$. Most of the 
contribution due to point sources is removed. Strong diffuse Fe K$\alpha$ emission 
is present close to the Galactic plane east of Sgr A*. The green ellipses at low Galactic 
latitude show the background extraction regions. Galactic North is up.}
\label{FeK}
\end{figure*}

\begin{figure*}
\includegraphics[width=0.92\textwidth,height=0.685\textwidth,angle=0]{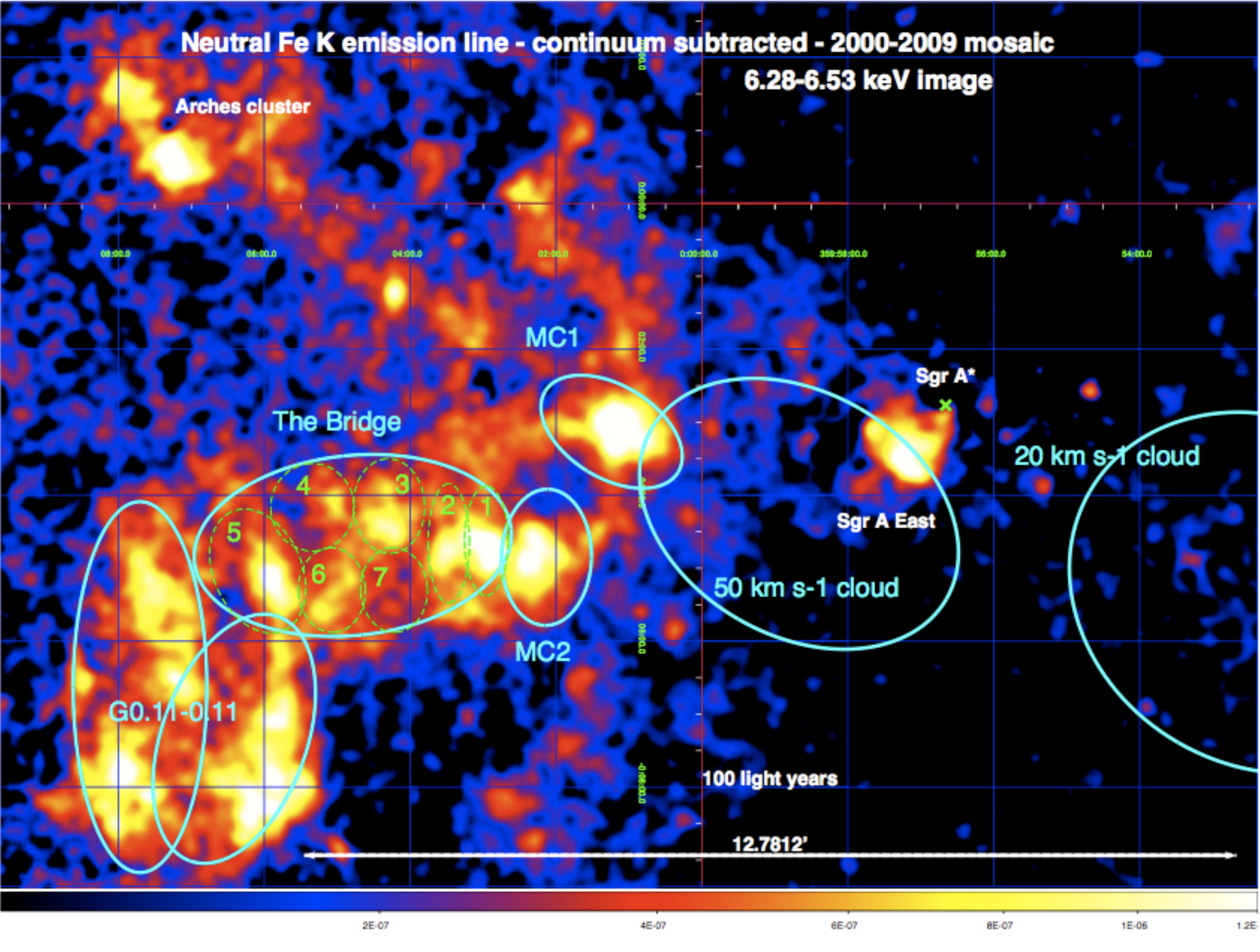}
\caption{Enlargement of the Fe K, continuum subtracted intensity map.
The solid light blue ellipses show the regions selected through the study 
of the CS maps and the comparison with the Fe K map. These should 
indicate and separate the emission from the different molecular clouds.
The corresponding background regions have been selected from an area 
at high latitude, south of the GC (see right panel of Fig. \ref{FeK}).
The green dashed ellipses show the subdivision of the bridge area 
into smaller regions.}
\label{FeK_EW}
\end{figure*}

The left panel of Figure \ref{FeK} shows an image of the entire EPIC field
of view in the 6.28-6.53 keV energy band. 
This band contains the K$\alpha$ fluorescence line of the neutral 
or low-ionisation Fe atoms, that peaks at 6.4 keV.
The image is the result of the mosaic of the images of each observation 
and the three cameras aboard \xmm.
The stronger features are due to transient point sources with strong 
continua and with only weak 6.4 keV lines. To subtract the emission 
from these sources and to disentangle the emission due to the 
Fe K$\alpha$ line, we first computed the continuum emission image 
in the 4.5-6.28 keV band, then assumed that the continuum spectral 
shape is described by an absorbed power law shape with spectral 
index $\Gamma=2$. This allowed us to calculate the continuum contribution 
underlying the Fe K line emission and thus subtract it after properly 
re-scaling the continuum image (a factor of 0.106 has been used). 
The right panel of Fig. \ref{FeK} shows that, once the continuum is 
properly subtracted, the emission from AX J1745.6-2901, GRS 1741.9-2853,
XMM J174554.4-285456, the Arches cluster, the cannonball, 
CXO J174621-285257 and the other point sources is correctly subtracted.
Only Sgr A East leaves a clear residual, because of its strong 6.7 keV line, 
contaminating also the Fe K$\alpha$ region with its wings.

If the latter region is ignored, one can easily see that a strong, diffuse, 
Fe K$\alpha$ emission at $\sim$6.4 keV is present in the region, and that 
it is distributed along the Galactic plane in a non-uniform way with several 
hot spots, more numerous and prominent on the galactic east side of Sgr A*.
Figure \ref{FeK_EW} shows an enlargement of this region.
As already well known, a similar general asymmetry is present in the 
molecular cloud distribution, suggesting a strong link between the 6.4 keV 
hot spots and MC complexes.
We therefore perform an accurate study of the correlation between 
the bright 6.4 keV line structures and the molecular matter distribution, 
using public data of molecular line emission of the GC region.

\subsection{Comparison with molecular CS emission}

\begin{figure*}
\includegraphics[width=0.92\textwidth,height=0.58\textwidth,angle=0]{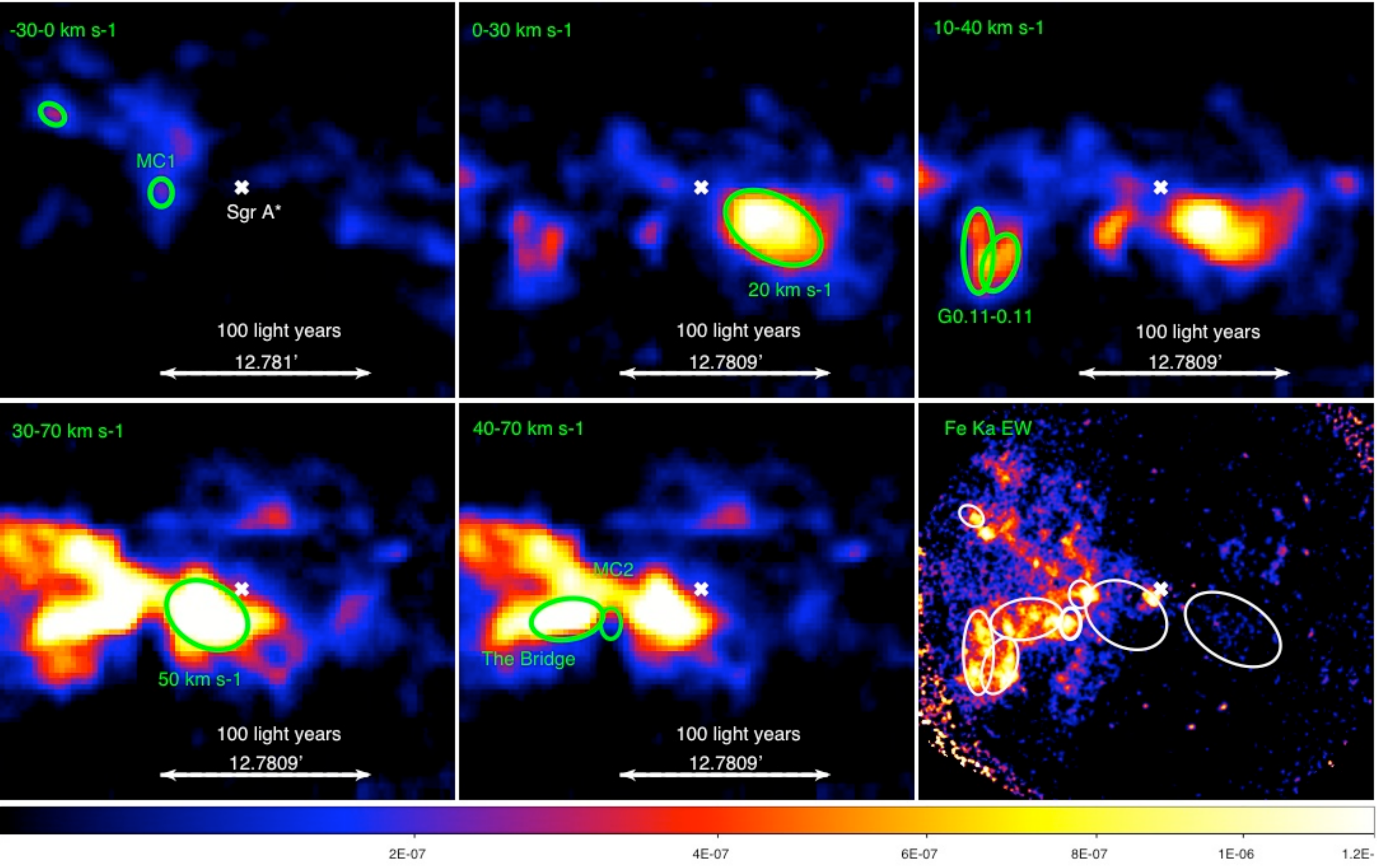}
\caption{CS maps from (Tsuboi et al. 1999). The different images are integrated 
over different velocity ranges to differentiate the coherent structures and try to 
separate emission coming from different regions along the line-of-sight.
{\it (Lower right panel)} Continuum subtracted Fe K$\alpha$ image. }
\label{CS}
\end{figure*}
The molecular gas of the Galaxy is strongly concentrated within the central few 
hundred light years (Bally et al. 1987; Tsuboi et al. 1999) in the so-called 
Central Molecular Zone (CMZ). About 10~\% of the molecular gas of the Galaxy is 
contained in this region (the CMZ has an estimated mass of several $\times$10$^7$ 
M$_{\odot}$) and it is less than 1 \% of the Galaxy volume. 
MC in the CMZ are known to have a higher density relative to the Galactic 
norm and for these reasons they are better depicted through high-density 
tracers like CS (Bally, et al. 1987; Tsuboi et al. 1999), 
NH$_3$ (Gusten et al. 1981; Morris et al. 1983) and HCN (Fukui et al. 1977; 
Jackson et al. 1996) lines than the CO line, usually employed for the rest 
of the Galaxy.
To map the location of the more dense and compact MC, we use the CS J=1-0 
data-cube provided by Tsuboi et al. (1999). Large velocities gradients are 
experienced by the molecular gas within the CMZ. Most probably, the 
clouds are not in Keplerian motion around the central BH. In this region, 
in fact, the mass concentration dominating the gravitational potential is 
given by stars that appear to be distributed in a bar. 
Because of the resulting non-circular motions, it is not possible to uniquely 
infer the distance to a MC from its position in the sky and its velocity along the 
line-of-sight. 

Several authors have attempted to provide 3D distribution of the molecular material
in the CMZ using different tools to reconstruct the MC position along the line-of-sight.
Vollmer et al. (2003) observed a population of infra-red point sources in the 
Galactic centre and, assuming an axi-symmetric distribution, they 
estimated the MC position through their observed extinction
(see also Becklin et al. 1982; Glass et al. 1987). Nevertheless, this technique 
provides only a very rough estimate of the distance. 
Up to now it has been possible to apply parallax measurements only to Sgr B2 
(Reid et al. 2009). Sawada et al (2004) proposed a method based on the 
comparison between emission and absorption in the radio band and applied it 
to the entire CMZ. Ryu et al. (2009) proposed a method based on the X-ray 
absorption toward the Sgr B region. Nevertheless, these interesting studies 
do not have enough resolution to measure the line-of-sight position of a 
single MC within the CMZ. 
On the other hand, single or spatially connected MC structures should have similar 
velocities. Selecting the MC emission in different velocity ranges, it might be possible 
to separate the MC distribution along the line-of-sight. 

The different panels of Figure \ref{CS} show the CS sky images 
corresponding to the EPIC-pn field (lower-right panel) integrated over 
different velocity ranges, in order to differentiate the various MC.
The intensity of CS emission (directly correlated with the MC mass 
when, as in the present case, CS auto-absorption does not occur) 
is shown in Fig. \ref{CS} with the same linear colour scale going 
from T$_{MB}=$2.85 K km s$^{-1}$ (in order to select only massive 
MC) to 60 K km s$^{-1}$. 
Using the formula from Tsuboi et al. (1999): 
n$_H = \frac{7.5 \times 10^{11} \times T_{ex} \times 
{\int T_{MB}dv} }{10^{-8} }$ cm$^{-2}$; we derive, 
for those values, column densities equal to n$_H$=4.3$\times10^{21}$ and 
9$\times10^{22}$ cm$^{-1}$, respectively (assuming abundance 
X(CS)=$10^{-8}$; Irvine, Goldsmith \& Hjalmarson 1987; and excitation 
temperature of CS of T$_{ex}=20$ K\footnote{Tsuboi et al. (1999) assume 
in their calculations an excitation 
temperature between 30 and 70 K, following the temperature 
measurements of the MC in the GC (Gusten et al. 1981; Morris et al.
1983; Armstrong et al. 1985; G\"usten 1989). Nevertheless, several 
authors studying ammonia and SiO lines show that these lines are 
sub-thermally excited for typical MC within the CMZ (Gusten et al. 1981; 
Morris et al. 1983; Amo-Baladron et al. 2009). In particular, Amo-Baladron 
et al. (2009) calculate the column densities of several MC in 
Sgr A, through the detailed measurement of the SiO excitation temperature,
performed with a LVG excitation code. From these values it is possible 
to deduce the equivalent CS excitation temperature T$_{ex}$ that results 
to be between 10 and 15 K. We thus assume a T$_{ex}$ of 20 K, a compromise 
between these values and the minimum value considered by 
Tsuboi et al. (1999).}). 
Table 2 shows the physical parameters of the different MC that we identified 
as measured from the X-ray and CS data. The second column shows the 
column density (n$_H$), the third the mean radius (R), the fourth the projected 
distance from Sgr A* (D$_p$) and the fifth the line-of-sight distance with 
respect to the plane of Sgr A* (D$_{los}$). To calculate the MC column 
densities from the CS maps we summed the CS emission in the -30--0, 
0--30, 30--70 and 40--70 km \s\ velocity ranges for the MC1, 20 km \s, 
50 km \s\ and the bridge MC, respectively.
No major massive MC is observed characterised by either velocities lower 
than -30 km s$^{-1}$, or higher than 70 km s$^{-1}$, thus we do not show 
the corresponding images. 

Figure \ref{CS} shows that the same asymmetry present 
in the Fe K line intensity image is present also in the MC distribution. 
Moreover, the more massive MC in the EPIC field of view are the 20 and 50 km 
s$^{-1}$ MC. They are located within 10-20 pc from the 
supermassive BH (Coil et al. 2000), their position with respect to Sgr A* is 
well known because of the physical interactions between these MC and Sgr 
East. The two MC have velocities between 0--30 km \s\ and 30--70 km \s,
respectively (see middle--upper and left--lower panels of Fig. \ref{CS}). 

Another well studied and massive MC in the field of view is G0.11-0.11 (Handa 
et al. 2006; Amo-Baladron et al. 2009). It appears in the velocity range between 
10 and 40 km \s\ (see upper right panel of Fig. \ref{CS}). Its shape is not easily 
reproducible by a single ellipsoid, thus we combine two ellipses to define it.

Two other bright Fe K spots appear in the X-ray image. 
Muno et al. (2007), analysing the \chandra\ data (see the green square in 
the left panel of Fig. \ref{FeK} that corresponds to the region studied by 
Muno et al. 2007), already studied these two Fe K bright regions and 
indicated as feature 1 and 2 (see Fig. 2 of Muno et al. 2007).
We keep the same name calling them MC1 and MC2.
Scanning the CS maps, we find a possible counterpart for MC1
in the velocity range between -30 and 0 km \s\ (see upper--left panel of 
Fig. \ref{CS}), while we do not see a clear counterpart for MC2. 
For this reason we can place only an upper limit to its column density 
(see Tab. \ref{MC_par}).
The ammonia and CO emission from MC1 has been studied by
Armstrong et al. (1985), calling it M0.02-0.05. The association of 
MC1 with the CMZ is still unclear. In fact, although the column 
density and temperature of this feature are typical of the CMZ,
Amo-Baladron et al. 2009 show SiO velocities for this 
cloud between -12.5 and -22.5 km \s. These velocities are consistent 
with the ones observed in the Galactic disk, but the linewidths seem
to be broad, characteristic of the GC.

In the range between 40 and 70 km s$^{-1}$ (see lower--middle
panel of Fig. \ref{CS}) an intense structure appear in the CS maps with a 
reversed V shape. The horizontal branch spatially corresponds to strong 
low-ionisation Fe K emission, while weaker emission is measured at the 
position of the vertical branch (see the lower--right panel of Fig. \ref{CS}).
The horizontal branch appears just east of the location of the 50 km 
s$^{-1}$ MC and reaches the location of G0.11-0.11 (the branch and 
G0.11-0.11 appear in different velocity ranges, suggesting that these 
are not interacting, being at different distances along the line-of-sight). 
This feature appears like a unique structure of molecular gas, clearly 
more extended in the longitudinal direction. For this reason and in order 
not to confuse the reader with terms already used to indicate other famous 
structures in the GC, we call it (as Armstrong et al. 1985 and Sakano et 
al. 2006) "the bridge". 
We note that in Fig. \ref{FeK_EW} the bridge seems to further extend 
to the west without discontinuity into region MC2. Nevertheless 
the CS maps show a clear transition with the bridge ending before 
MC2. Thus the bridge and MC2 seem to be two separate MC structures.

We also mention that a bright spot in the upper left side of the X-ray image 
has a molecular counterpart. This region is associated to the Arches cluster 
and has been already studied using \xmm\ and \chandra\ data 
(Yusef-Zadeh et al. 2002; Sakano et al. 2004; Wang et al. 2006). We will not, 
therefore, discuss this feature further in this paper. 

\begin{table}
\small
\begin{center}
\begin{tabular}{l cc cc cc}
\hline 
\hline 
name                & n$_H$                            &  R   & D$_p$ & D$_{los}$ & $\Delta$ V$_{\rm LSR}$\\
                         & (cm$^{-2}$)                    & (pc) & (pc)    & (pc) & km \s \\
\hline
Bridge                  &9$\times10^{22}\dag$   &1.6   & 18     & 60\ddag & 40--70 \\
G0.11-0            & 2$\times$10$^{22}$\dag\dag& 3.7  &  25     & 17$\ddag$  & 10--40 \\
50 km s$^{-1}$&9$\times10^{22}\dag$    & 4   & 5         & 5-10 & 30--70 \\
MC1                 &4$\times10^{22}\dag$   & 1.8 & 12    &50\ddag & -30--0 \\
MC2                 &$<$2$\times10^{22}\dag$   &1.8  & 14       & $<$35\ddag & \\
\hline
Sgr B2              & 8$\times$10$^{23}$         & 7 & 100   & -130\ddag\ddag & \\
\hline
\hline
\end{tabular}
\end{center}
\caption{MC physical parameters: 
MC column density (n$_H$); radius (R);
projected distance from Sgr A* (D$_p$); distance along the line-of-sight 
(D$_{los}$) with respect to the plane of Sgr A* (negative values indicate 
objects closer to us than Sgr A*); CS integration velocity range ($\Delta$ V$_{\rm LSR}$).
$\dag$ Column density inferred from the 
CS maps of Tsuboi et al (1999); $\ddag$ value estimated assuming that 
the MC is illuminated by the same flare illuminating Sgr B2; \dag\dag 
Amo-Baladron et al. 2009; \ddag\ddag Reid et al. 2009. }
\label{MC_par}
\end{table} 

In summary, the comparison between Fe K and CS emission shows a strong 
correlation between the two. In particular we observe that MC emission 
is present wherever Fe K$\alpha$ emission is bright; on the other hand, 
massive molecular complexes are present without associated Fe K emission.
The solid green ellipses in Fig. \ref{FeK_EW} shows 
the selected regions from which the X-ray spectra will be extracted. The bigger 
region (the so called "bridge") has also been subdivided into further, smaller 
areas (shown in dashed light blue in Fig. \ref{FeK_EW}) that will be used 
to perform time-resolved spectral analysis. We also note that Fig. \ref{FeK_EW} 
shows other regions of diffuse Fe K emission. 
A more detailed study of this fainter component and its association to MC, is 
deferred to a later work. 

\section{Average spectral characteristic of MC}

\begin{table*}
\footnotesize
\begin{tabular}{l cc cc cc cc cc}
\hline 
\multicolumn{7}{l}{Model: wabs $\times$ ( apec + edge $\times$ ( power-law + Gaus + Gaus ) )} \\ \\
\hline 
name &
nh &
 $\tau$ &
E &
 $\sigma$ &
norm$_{Ga}$ &
 $\Gamma$ & 
norm$_{pl}$ &
EW &
$\chi^2$/dof \\
 &
10$^{22}$ &
 &
 &
 &
10$^{-5}$ &
&
10$^{-5}$ &
&
\\
 &
(cm$^{-2}$) &
&
(keV) &
(eV) &
(ph. cm$^{-2}$ s$^{-1}$) &
&
(ph. keV$^{-1}$ cm$^{-2}$ s$^{-1}$) &
(eV) &
\\

Bridge &
4$\pm3$ &
0.26$\pm0.12$ &
6.409$\pm$0.002${\dag}$ &
28$\pm$4$\dag\dag$ &
4.7$^{+0.3}_{-0.2}$ &
1.0$^{+0.4}_{-0.3}$ &
26$^{+22}_{-13}$ &
750 &
1175.1/1121 \\

G0.11-0.11 &
7$\pm4$ &
0.03$^{+0.11}_{-0.03}$ &
6.411$\pm$0.002 &
28$\pm$5 &
7.5$\pm$0.5 &
1.9$^{+0.3}_{-0.4}$ &
250$^{+200}_{-130}$ &
955 &
1302.0/1175 \\

MC1 &
10$^{+1}_{-2}$ &
0.32$\pm$0.07 &
6.410$\pm$0.005 &
$<18$ &
1.87$^{+0.18}_{-0.06}$ &
0.8$^{+0.4}_{-0.5}$ &
10$^{+10}_{-5}$ &
684 &
780.3/780 \\

MC2 &
5$^{+5}_{-4}$ &
0.36$^{+0.2}_{-0.15}$ &
6.411$\pm$0.004 &
30$^{+7}_{-10}$ &
0.98$^{+0.17}_{-0.09}$ &
0.9$^{+0.1}_{-0.5}$ &
6.3$^{+0.5}_{-2}$ &
715 &
755.1/615 \\
\hline
\hline
\multicolumn{7}{l}{Model: wabs $\times$ (power-law + apec + pexrav+ Gaus + Gaus)} \\ \\
\hline 
name &
nh &
 &
E &
$\sigma$ &
norm$_{Ga}$ &
$\Gamma$ & 
norm$_{pl}$ &
&
$\chi^2$/dof \\

Bridge &
4$\pm$3 &
&
6.409$\pm$0.002 &
26$\pm$5 &
4.6$\pm$0.2 &
2.1$\pm0.1$ &
78$^{+55}_{-34}$ &
&
1179.7/1122 \\

G0.11-0.11 &
6$^{+2}_{-3}$ &
&
6.411$\pm$0.002 &
27$\pm$5 &
7.1$^{+0.4}_{-0.3}$ &
2.4$\pm$0.2 &
420$^{+160}_{-130}$ &
&
1309.5/1176\\

MC1 &
9$^{+2}_{-4}$ &
&
6.410$\pm$0.005 &
$<20$ &
1.9$\pm$0.07 &
1.9$\pm0.1$ &
28$^{+24}_{-15}$ &
&
781.8/781 \\

MC2 &
8$^{+6}_{-5}$ &
&
6.409$\pm$0.004 &
23$\pm$10 &
1.2$\pm$0.1 &
2.1$\pm$0.2 &
35$^{+18}_{-10}$ &
&
760.0/615\\
\hline 
\hline 
\end{tabular}
\caption{\rm Best fit results of the fit of the mean spectra of the different MC.
EPIC-pn and MOS data are fitted simultaneously. {\it (Upper panel)} 
Fit with the phenomenological model. {\it (Lower panel)} Fit including the 
reflection continuum.
Column densities are in units of 10$^{22}$ \cm;
Fe K$\alpha$ normalisation in units of 10$^{-5}$ ph \cm \s;
power law normalisation in units of 10$^{-5}$ ph. keV$^{-1}$ cm$^{-2}$ \s\ 
at 1 keV. \dag Please note that 
the calibration of the absolute energy scale of the EPIC instruments is 
known within 10 eV uncertainty (CAL-TN-0018), higher than the Fe K$\alpha$ 
statistical uncertainty ; \dag\dag The EPIC-pn and MOS energy resolution at 
the Fe K energy is of the order of $\sigma$$\sim$60 eV.}
\label{fit}
\end{table*} 

\begin{figure}
\includegraphics[width=0.34\textwidth,height=0.46\textwidth,angle=-90]{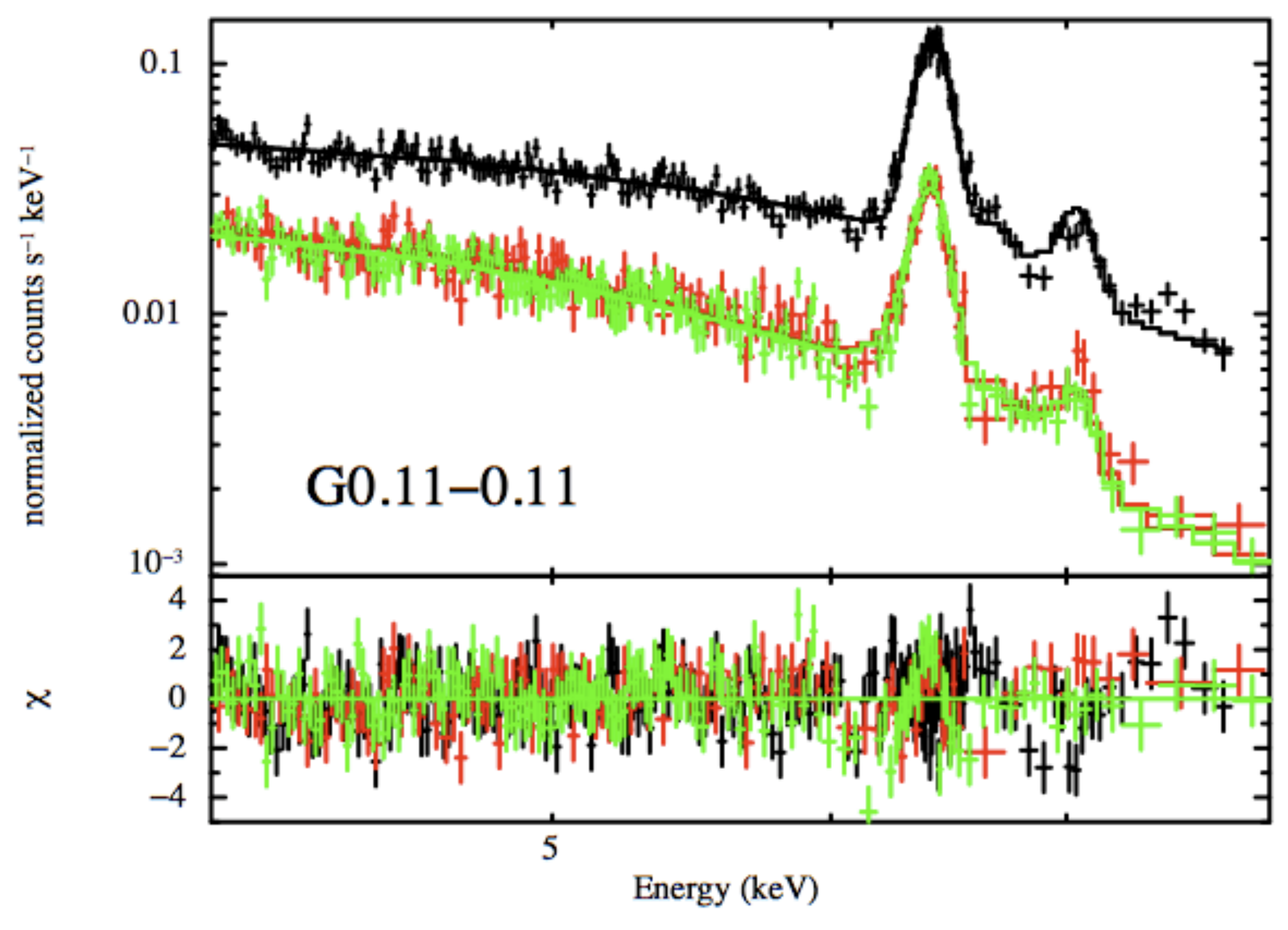}
\includegraphics[width=0.34\textwidth,height=0.46\textwidth,angle=-90]{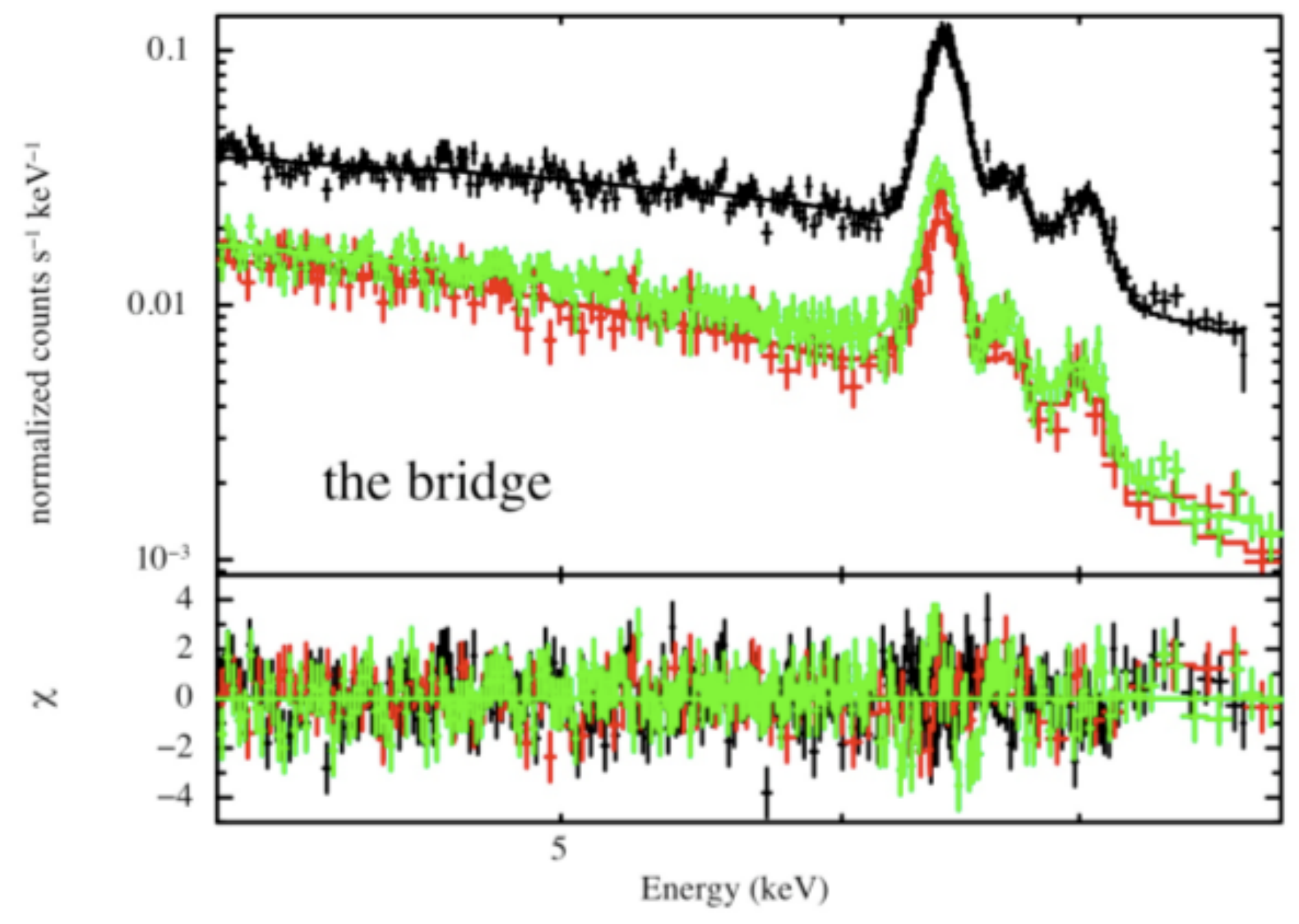}
\caption{X-ray spectrum of G0.11-0.11 and the bridge. EPIC-pn in black, MOS1 in red 
and MOS2 in green. Clearly evident are the Fe K$\alpha$+$\beta$ lines, 
the Fe K edge, the flat power law continuum.}
\label{specG011}
\end{figure}

We compute the spectra of the different MC regions, delimited by the 
solid green ellipses in Fig. \ref{FeK_EW}. 
We then add all the spectra and response 
files with the {\sc ftool} {\sc mathpha}, keeping the different instruments separated. 
We use local background spectra (with the same extraction area as for the respective 
MC region) from positions at high Galactic latitude South of Sgr A*, within the 
same field of view (see right panel of Fig. \ref{FeK}). The subtraction of local 
background spectra almost completely removes the contribution due to 
the hard component of the X-ray diffuse emission (see Fig. \ref{specG011}), 
including the 6.7 keV line (Revnivtsev et al. 2009), this latter being almost 
completely uniform on these angular scales. It also removes part of the 
emission from the low temperature (about 1 keV) plasma.
However, residual soft lines appear in the X-ray spectra of the different MC. 
For this reason we present the results of the spectral analysis considering 
only the 4-8 keV band.
However, similar results are obtained considering the entire 
band (2-10 keV), once the contribution due to a low temperature plasma 
is considered to fit the low energy band.

As examples of MC emission, Fig. \ref{specG011} shows the X-ray 
spectrum of G0.11-0.11 and the bridge. 
The strongest spectral features are: i) the neutral-low ionised Fe K$\alpha$ 
line and the associated Fe K$\beta$; ii) a power-law continuum absorbed by 
neutral material; iii) an Fe K edge; iv) a residual ionised Fe K emission (the line 
is almost completely removed with the subtraction of the local background). 
We thus fit the data with  a phenomenological model given by the
{\sc Xspec} model: 
{\sc wabs*(apec+edge*(power-law+Gaus+Gaus))}.
Table \ref{fit} shows the best fit results of the mean spectra of the different MC. 
The Fe K edge energy is fixed at 7.1 keV, the Fe K $\beta$ emission energy fixed at 
7.06 keV (as in Ponti et al. 2009), its width fixed 
to that of Fe K$\alpha$ and its intensity tied to that of Fe K$\alpha$ in 
order to keep a ratio $\beta$/$\alpha$=0.15. We also assume Solar abundance 
(Grevesse \& Anders 1989) for the emitting material and fix the temperature of 
the thermal ({\sc apec}) component to 6.5 keV (Koyama et al. 2007). 

We observe (see Tab. \ref{fit}) that the line energy is, in about all MC, 
E=$6.410\pm0.002$ keV, a value higher than the expected Fe K$\alpha$ 
emission from neutral iron; nevertheless once the systematic uncertainties are 
considered, it becomes consistent with emission from neutral Fe\footnote{The 
observed energy corresponds to emission from Fe {\sc xii-xiv} (House 1969) 
that, assuming photo-ionisation, would correspond to an ionisation factor 
of about log($\xi$)=1-1.5 erg cm \s\ (Makishima 1986). 
Nevertheless, we note that the calibration of the absolute energy scale 
of the EPIC instruments is known within 10 eV (CAL-TN-0018), thus the observed 
energy is still consistent with emission from neutral iron. The high value of the 
energy of the Fe K emission might be induced by the periods of high soft 
proton activities occurring during the observation (see Tab. 1). 
In fact, when the MOS data alone are considered (that have absolute energy
scale uncertainty of 5 eV only) the Fe K energy results to be 
E$_{FeK}=6.405\pm0.005$ keV.}.
Also the line width is statistically not consistent with 0, being of the order 
of $\sigma\sim20-30$ eV. However the lines are not significantly broad 
when the systematic uncertainties are considered (Guainazzi et al. 2010)\footnote{
Fitting simultaneously EPIC-pn and MOS data, we obtain line widths of the order 
of 20-30 eV. On the other hand, upper limits of 9 eV are obtained when the MOS 
data only are considered, while a width of $39\pm5$ eV is obtained fitting 
EPIC-pn data only. Moreover, following Guainazzi et al. (2010) we checked 
that (analysing {\sc calclosed} observations) the calibration lines measured by 
the EPIC-pn camera have systematic width of the order of 20-40 eV. 
The measured Fe K lines are thus consistent with being narrow.}.

The Fe K line 
has equivalent width (EW) between $\sim$0.7 and 1 keV above the total observed 
continuum. Such high EW are in agreement with a reflection origin for 
the line and they imply a minor contribution from other sources of 
continuum different from reflection. The power law continuum is observed 
to be extremely flat with a spectral index between about $\Gamma\sim$0.6 
and 1.7. In the majority of the MC, the Fe K edge is detected with high 
significance (between 65 and 99.9 \%). 

We thus add to the model a component reproducing a reflection continuum 
associated with the Fe K line (the {\sc pexrav} model in {\sc Xspec}). 
The illuminating power law is assumed to have a spectral index 
$\Gamma$=2 with a high energy cut-off at 150 keV. Solar abundance and 
60 degrees inclination are assumed for the reflecting material. 
Moreover, the reflection continuum 
normalisation is fixed such as to produce a Fe K line of about 1 keV above the 
reflection continuum. The major features of a reflection component, from cold 
irradiated material with significant optical depth, are a low ionisation Fe K line
with equivalent width of about 1 keV, a pronounced Fe K edge feature, a 
Compton hump around 20--30 keV, and a flat 2--10 keV continuum 
($\Gamma\sim0$) because of the internal photo-absorption. 

Once the reflection continuum is added, we observe 
no significant changes in the quality of the fit (see lower panel of Tab. \ref{fit}). 
Nevertheless, the Fe K edge (being a predicted feature of the model) is not required 
anymore and the direct power law emission has a steeper spectral index, consistent 
now with the values observed in AGN (Ponti et al. 2006; Dadina et al. 2007; Bianchi 
et al. 2009) and accreting Galactic sources (McClintock \& Remillard 2006).
Fe K bright MC seem, thus, consistent with having a X-ray emission spectrum 
dominated by a strong reflection component.

Within the inner 15 arcmin from Sgr A*, massive molecular clouds without 
significant neutral Fe K$\alpha$ emission are also present (see Fig. \ref{FeK_EW}). 
The spectrum of the 50 km s$^{-1}$ cloud is dominated by the emission 
of Sgr A East, showing an intense (EW$\sim$1 keV) Fe K line at $\sim$6.66 keV. 
The line is relatively narrow ($\sigma=70\pm2$ eV). Nevertheless, 
it shows small red and blue wings. Once the red wing is fitted with a 
narrow neutral Fe K line, we measure a Fe K intensity of
1.65$\pm0.11\times10^{-5}$ ph cm$^{-2}$ s$^{-1}$ (EW=45 eV).
It is not clear if this component is either tied to the supernova remnant
or to the 50 km s$^{-1}$ cloud, for this reason we consider this as an upper 
limit on the emission from the 50 km \s\ cloud.

Also the 20 km \s\ cloud shows a prominent line at 6.66-6.7 keV. 
As in the 50 km \s\ cloud the neutral Fe K emission is weak with an 
intensity of about $1.2\pm0.1\times10^{-5}$ ph \cm \s\ and an EW 
of $\sim$60 eV.

\section{Time resolved spectral analysis}

\subsection{Variations of the Fe K line in the bridge region}

The high statistics in the Fe K band, associated with the good spatial resolution 
and the long monitoring provided by \xmm, allows the study of the Fe K line 
variations over the last 8 years of observations.
As shown in Table \ref{Obs}, several \xmm\ pointings have short exposures 
and are close in time. In order to improve the statistics during those periods 
we summed the spectra of different observations close in time, as indicated in 
Table \ref{Obs} which reports the grouping number for each observation. 
We, thus, obtained a significant measurement in 2002, one in March and one 
in November 2004, moreover one in April 2007, one in March 2008 and one 
in April 2009.

In order to have a measurement of the Fe K$\alpha$ intensity basically 
independent from continuum variations, we fit at each time the MC spectra 
with the phenomenological model composed of: 
{\sc wabs*(apec+edge*(power-law+Gaus+Gaus))}
instead of considering the more physical model including the reflection 
continuum component. 
We tested, however, that the same results can be obtained once  
the reflection continuum (tied to the Fe K line) is considered in the model.
The Fe K$\alpha$ line is, in fact, by far, the feature with the highest statistics
of the spectrum, thus its intensity does not depend on the other components 
of the model.

We first fit the spectra of the different instruments separately. We obtain 
consistent results for all observations, thus we fit the spectra of the three 
instruments simultaneously. 
In the fit we assume that the Fe abundance, the hot plasma temperature, 
the column density of the neutral absorber, the depth of the absorbing 
edge, and the width of the Fe K$\alpha$ line do not vary over time. 
We observe (as expected) that the intensity of the hot plasma emission 
is constant over time, thus we fix also this quantity.
\begin{figure}
\includegraphics[width=0.46\textwidth,height=0.34\textwidth,angle=0]{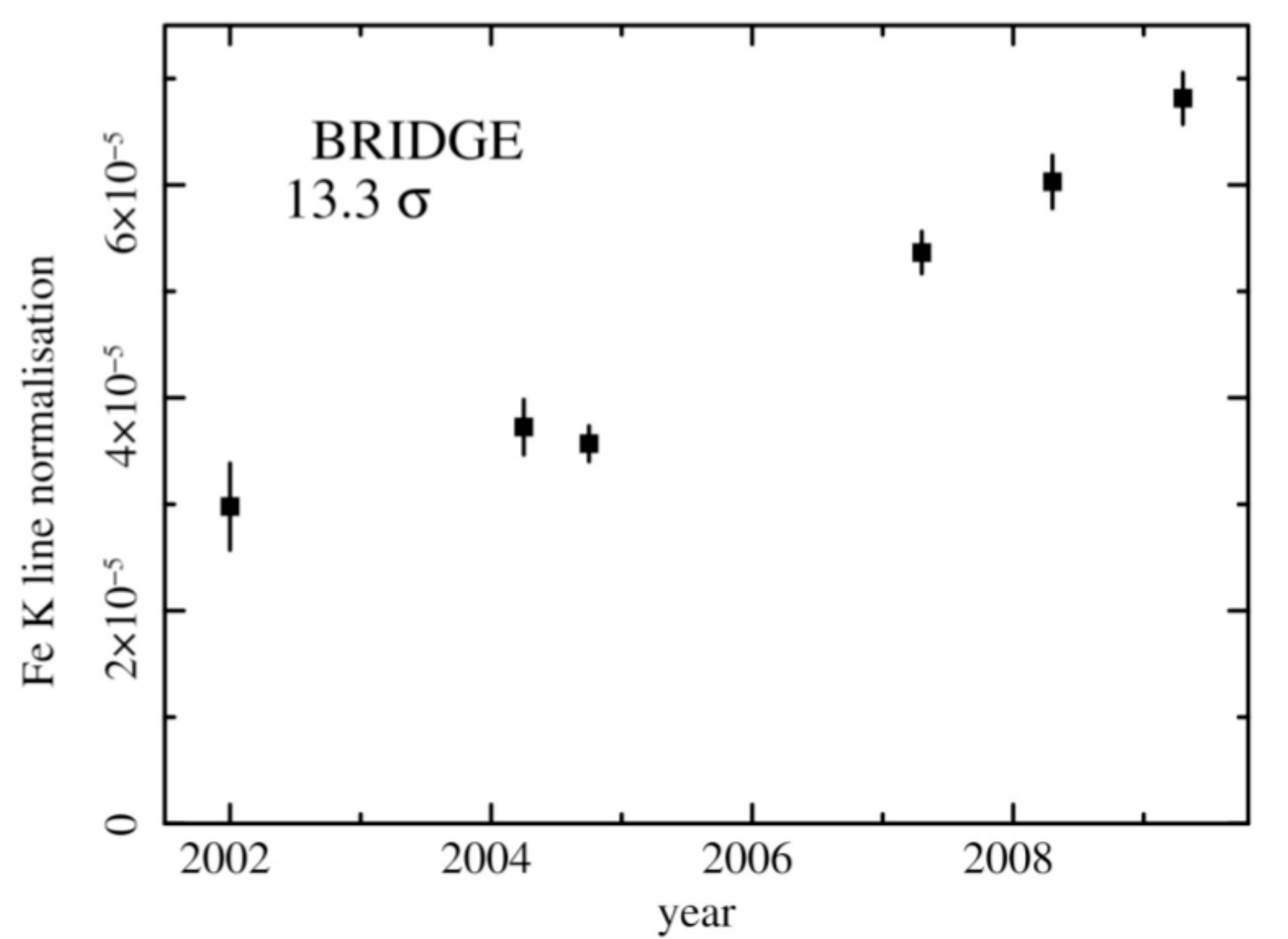}
\caption{Fe K intensity light curve of the bridge region. 
Extremely significant variations (at a level of about 13.3 $\sigma$) are occurring in 
this region. The main variations take place between 2006 and 2009.}
\label{lc_arm}
\end{figure}
\begin{figure*}
\includegraphics[width=0.6\textwidth,height=0.85\textwidth,angle=-90]{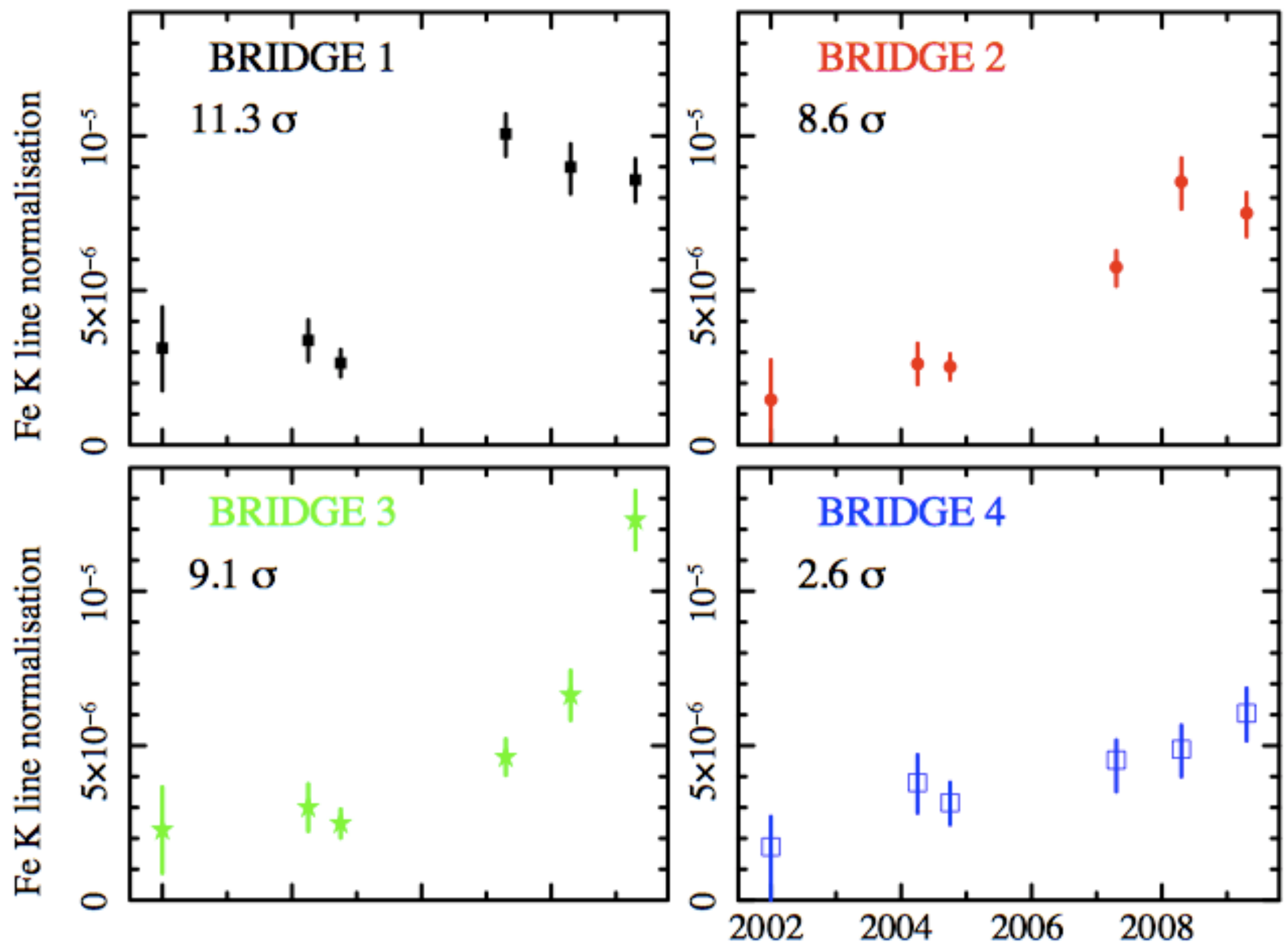}
\caption{
Fe K intensity light curves of the bridge 1, 2, 3 and 4 regions. 
Highly significant variations of the neutral Fe K line are seen in these regions.
The variations have significance of 11.3, 8.6 9.1 and 2.6 $\sigma$ for the 
bridge 1, 2, 3 and 4, respectively. In bridge 1 the peak of emission is reached between
2005 and 2007, in bridge 2 in 2008 and in bridge 3 in 2009. Surprisingly the different 
regions show similar amplitudes of variations (all about a factor of 3). 
The different light curves suggest they are undergoing a similar pattern of variations,
with, after a quiet period, a fast rise of reflection intensity in about 3 years (see bridge 3)
followed by either a stable or a decrease period of emission. In this scenario, bridge 4
might be just starting to rise.
}
\label{lc_arm14}
\end{figure*}
\begin{figure}
\includegraphics[width=0.46\textwidth,height=0.34\textwidth,angle=0]{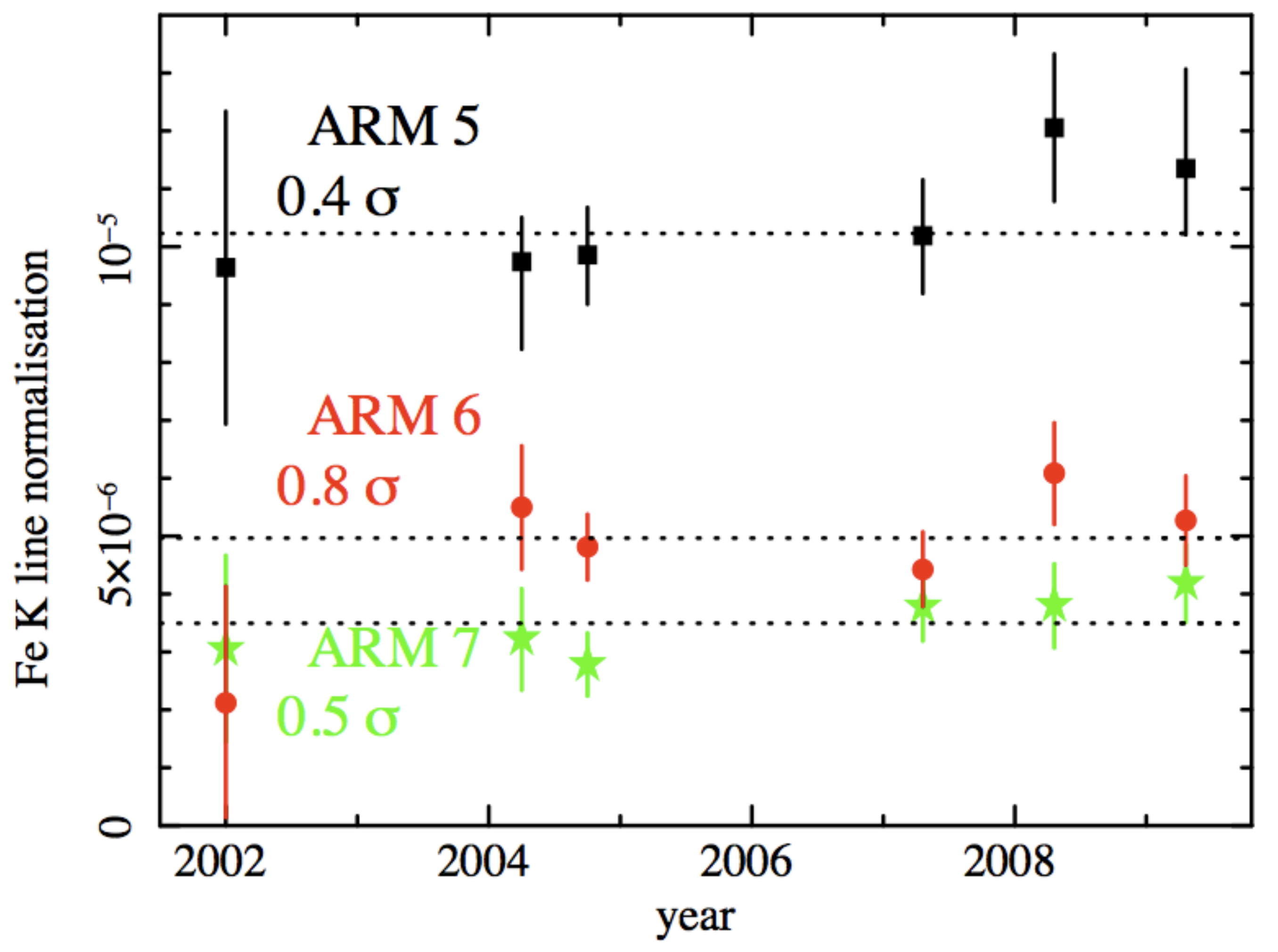}
\caption{Fe K intensity light curve of the bridge 5, 6 and 7 regions. The 
reflection component from these regions is consistent with being constant 
between 2002 and 2009. 
}
\label{lc_arm57}
\end{figure}

We also observe that the spectral index of the power law component 
is anti-correlated with the 4-10 keV flux. The spectral index becomes 
flatter at higher fluxes. This correlation is expected if the 4-10 keV 
continuum is composed by two components, a constant and steeper 
power law component and a flatter one due to a variable reflection.
Higher fluxes are thus associated with higher levels of reflection, thus 
flatter spectra. This interpretation is reinforced by the evidence that 
this correlation is not anymore significant once a reflection continuum tied 
to the Fe K line is added to the model. This suggests that the 
main 2-10 keV variations in these MC are due to the variability of 
the reflection component.

Figure \ref{lc_arm} shows the light curve of the intensity  
of the Fe K line in the bridge. This region experienced an increase 
of the Fe K line flux of a factor of about 2 in a few years.
The fit of the light curve with a constant gives a $\chi^2=196.2$ for 
5 dof, corresponding to a significance of the variation of 13.3 $\sigma$.
The region of the bridge is spatially quite extended (see Fig. \ref{FeK_EW}), 
thus, in order to better constrain where the variation occurs, we 
divide the bridge in smaller regions as indicated by the green dashed 
ellipses in Fig. \ref{FeK_EW}. We identify the sub-regions with a 
progressive number (starting from west and anti-clockwise) from 1 to 7. 

Figure \ref{lc_arm14} shows the light curve of the first 4 sub-regions, 
while Fig. \ref{lc_arm57} the variations of the regions from 5 to 7.
The fit with a constant gives a $\chi^2=145.8$, 89.9, 98.4, 15.4, 
3.1, 4.8 and 3.4 for 5 dof each, corresponding to a significance
of the variation of 11.3, 8.6, 9.1, 2.6, 0.41, 0.78 and 
0.46 $\sigma$, for bridge 1, 2, 3, 4, 5, 6 and 7, respectively.
Thus, the emission south of the bridge (in sub-regions 5, 6 and 7) is 
consistent with being constant, while highly significant variations of 
factors of about 3 are observed in the regions 1, 2 and 3. 
Interestingly the amplitude of the variations is similar in all 
the sub-regions from 1 to 3. We also note that all these regions 
(from bridge 1 to 4) have similar and significant line emission 
before the variation occurs, then not only a similar factor of 
variation, but also a similar intensity difference. Moreover, the peak 
of the variation occurs between 2005 and 2007 in bridge 1 (due to 
the lack of data is difficult to establish it with precision), while the 
peak is reached in 2008 in bridge 2 and in 2009 in bridge 3. 
This suggests a lag between the variation in the different regions.
This idea is reinforced by the similarity between the pattern of the 
variations in the three regions. A sharp rise of a factor of about 3 
in the intensity of the Fe K line occurs in about 3 years,
seen clearly in bridge 3 but consistent with bridge 2 and 1 trends.
After the peak, the Fe K line intensity light curve shows 
either a stable 
period, or a very slow decay (see bridge 1 and 2). Also the light 
curve of bridge 4 fits in this scheme, if this region is just entering 
in the steep rise phase. The similar shape of the variations suggests 
a deep connection between these variations. In particular it indicates 
a common mechanism producing such behaviours that might be tied 
to the propagation of some event either inside or outside the MC.
In this case, we foresee that the variations that occurred in the region
"bridge 1" should propagate to bridge 2, 3 and 4 in the next years.

\subsection{Variations of the Fe K line in other molecular structures}

\begin{figure}
\includegraphics[width=0.46\textwidth,height=0.34\textwidth,angle=0]{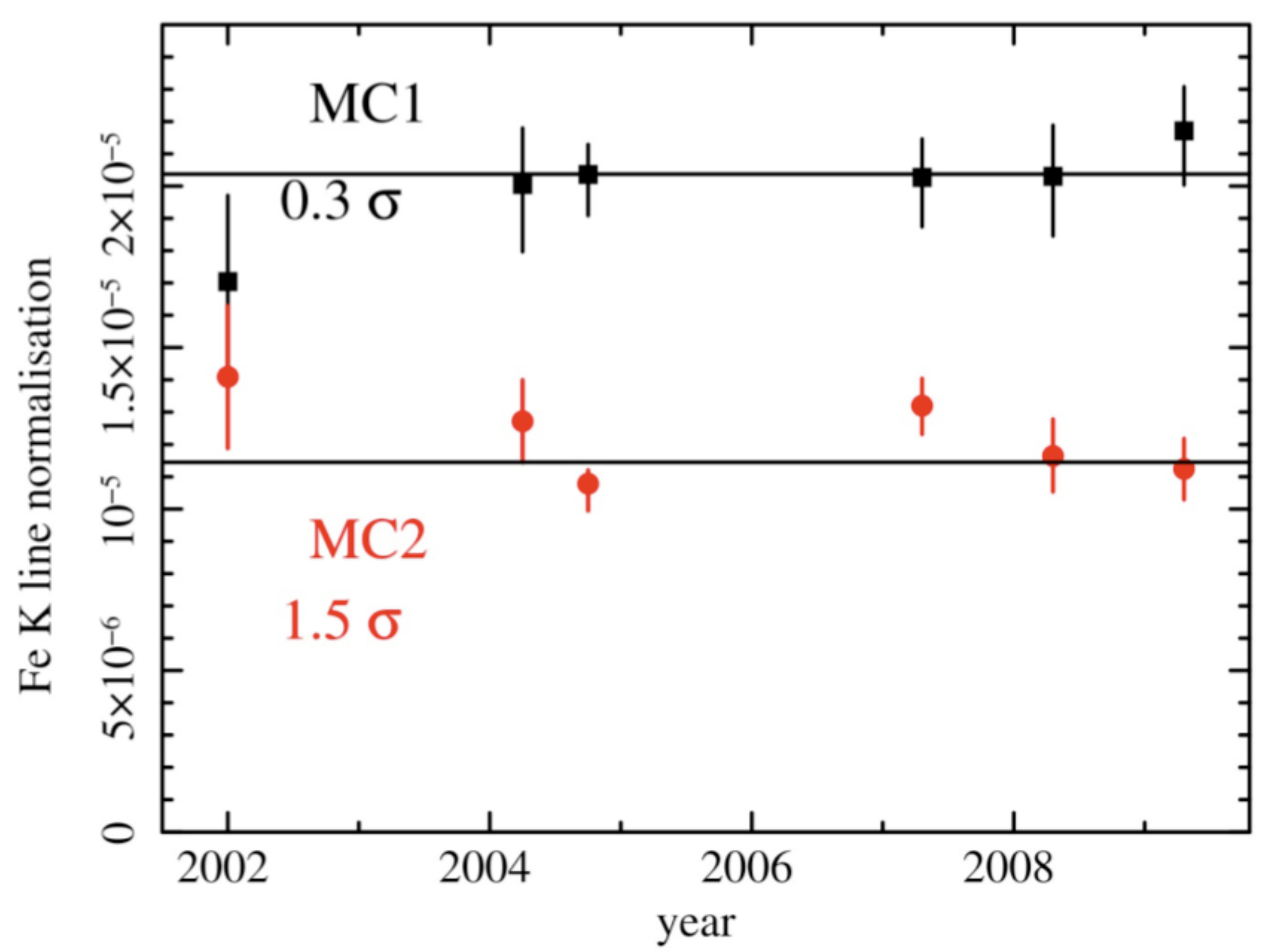}
\caption{Fe K intensity light curves of MC1 and MC2. The emission from these MC 
is consistent with being constant over the 8 years \xmm\ monitoring.}
\label{lc_mc12}
\end{figure}
Figure \ref{lc_mc12} shows the light curve of MC1 and MC2. 
Over the decade of \xmm\ observations, the Fe K line emission from these 
two MC is constant. In fact, the best fit with a constant is satisfactory, giving 
a $\chi^2=2.6$ and 9.4 for 5 dof each, corresponding to a significance of 
the variation of the order of 0.3 and 1.5 $\sigma$ for MC1 and MC2, 
respectively. We confirm the Muno et al. (2007) result of an overall 
constant Fe K emission coming from these MC and we note that the spatial 
variability observed with \chandra\ might not appear here because of the 
lower spatial resolution.
Unfortunately, the poor spatial resolution of the \suzaku\ telescopes does not 
permit a clear separation of the emission from MC1, MC2 and the bridge. 
The FeK line 
variability claimed by Koyama et al. (2009) to come from feature 1 of Muno et 
al. (2007) might, instead, be produced in the bridge region.

\begin{figure}
\includegraphics[width=0.46\textwidth,height=0.34\textwidth,angle=0]{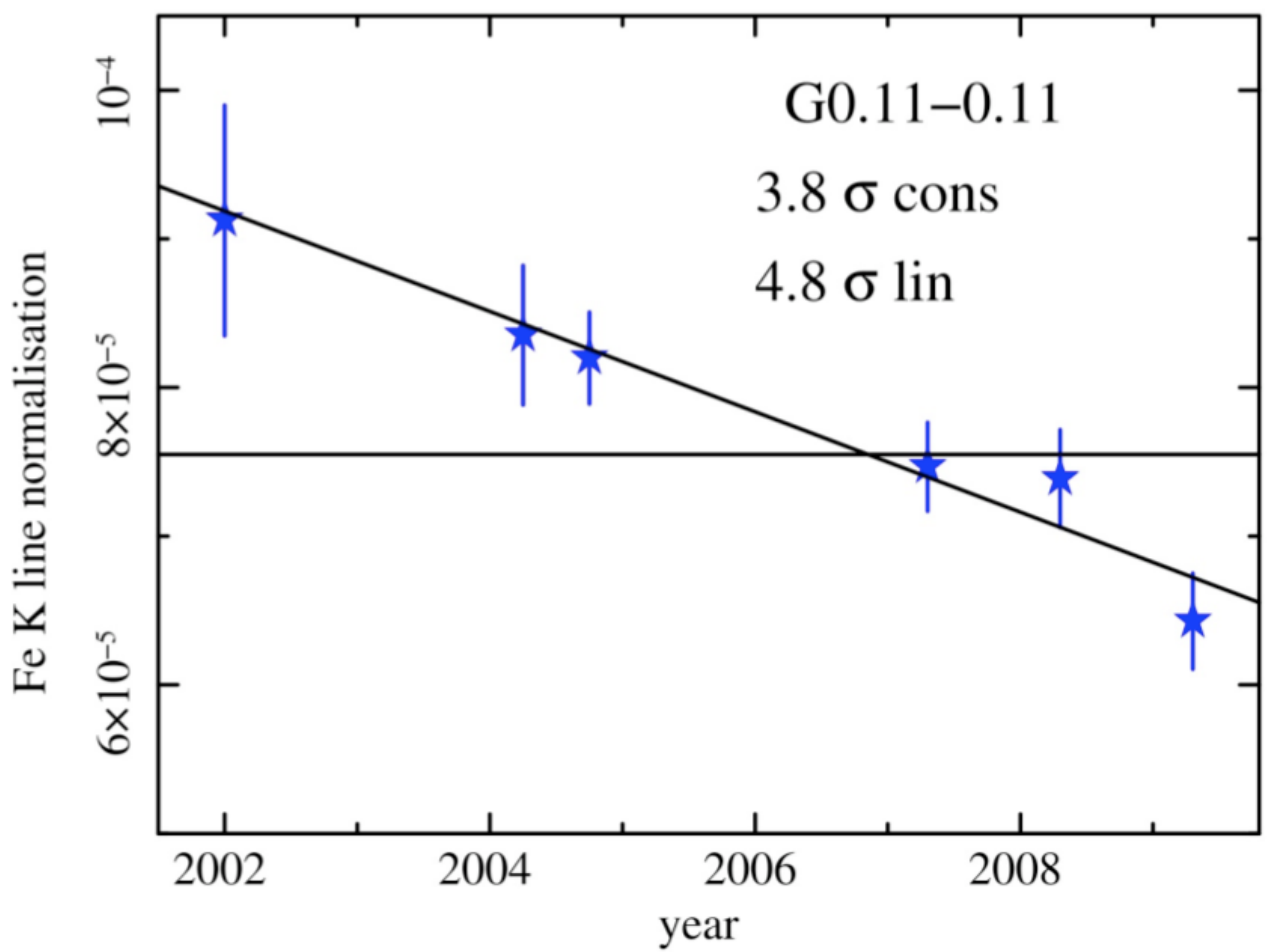}
\caption{Fe K intensity light curve of G0.11-0.11. The reflection features in this MC are 
clearly variable (at a significance level of about 3.8 $\sigma$) with a linear decay.
A linear decrease is a better fit at 4.8 $\sigma$ level. A similar decline is shown by 
the Sgr B2 MC. The two MC are consistent with reflecting the same Sgr A* flare 
(see text for details). }
\label{lc_G011}
\end{figure}
Figure \ref{lc_G011} shows the light curve of G0.11-0.11. 
The Fe K emission is clearly variable, with a strong decrease of the order 
of 50~\% within the 8 years of \xmm\ monitoring.
The fit with a constant gives a $\chi^2=24.2$ for 5 dof, corresponding to a 
significance of the variation of 3.8 $\sigma$. The light curve suggests a linear 
decrease. Fitting the data with a linear relation the fit improves, giving now a 
$\chi^2=1.9$ for 4 dof, corresponding to a significant improvement of about 
4.8 $\sigma$.

\section{Time resolved images:
discovery of a super-luminal echo in an X-ray reflection nebula}

If, as indicated by the time resolved spectral analysis, the variations in 
the different regions of the bridge are connected and due to the propagation
of an event inside the MC, the Fe K images between 2004 and 2009 should 
show the same variation and how it spatially had propagated. 
\begin{figure*}
\includegraphics[width=0.92\textwidth,height=0.74\textwidth,angle=0]{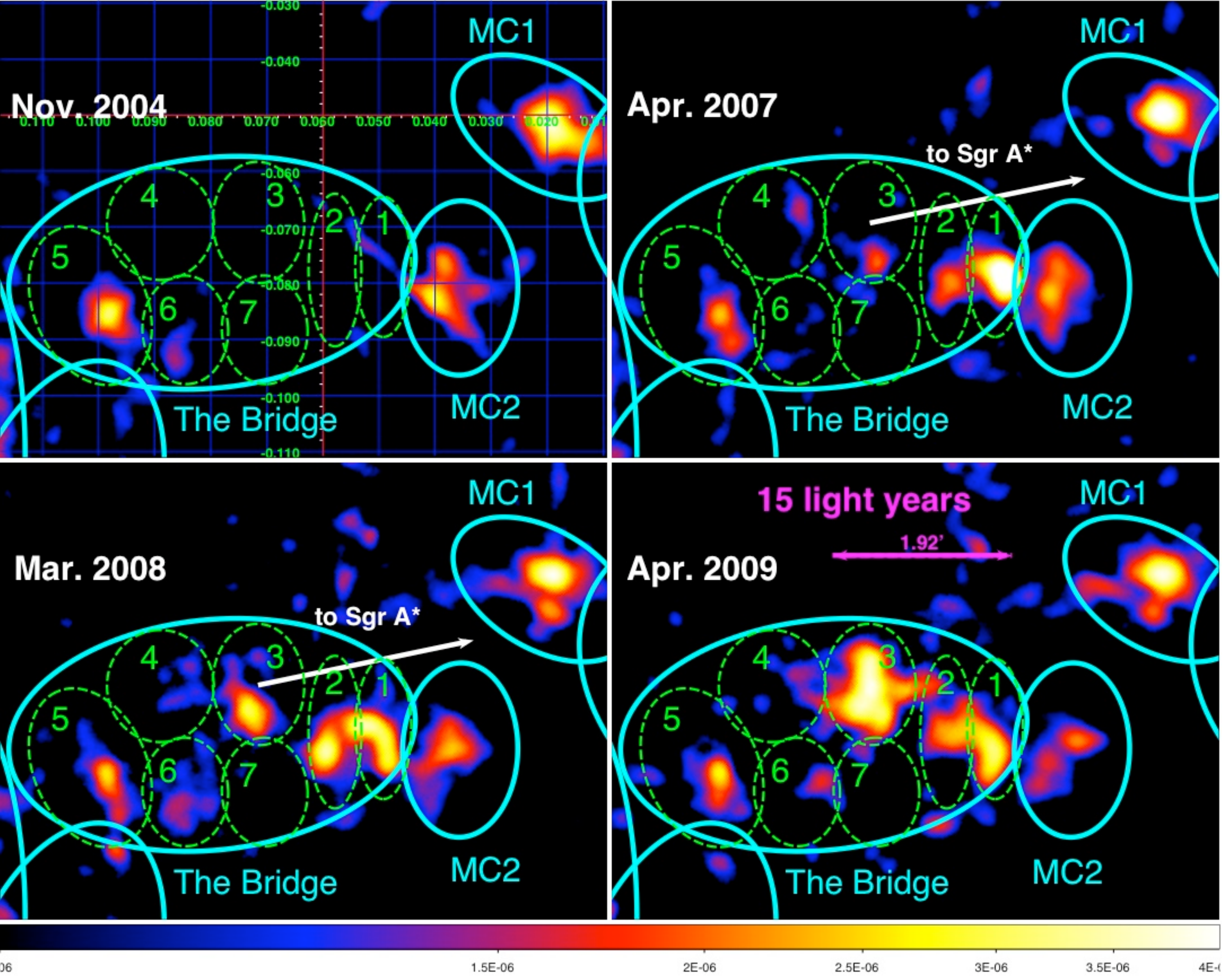}
\caption{Fe K$\alpha$ continuum subtracted mosaic image of the 
different EPIC-pn observations of the bridge region. A brightening of 
the bridge 1, 2, 3 and 4 is clear. Such variation occurs in a time-scale of 
about 2-4 years, but in a spatial scale of about 15 light years. 
This apparent super-luminal motion can be explained if the bridge 
MC is illuminated by a bright (L$>1.3\times10^{38}$ erg \s) and distant 
($>$15 pc) X-ray source active for several years. Either a flare 
from Sgr A* or a bright and long outburst of a X-ray binary can be 
the source of such a phenomenon. The observation of super-luminal 
echo can not be explained by either a single internal source or by low 
energy cosmic ray irradiation. It is also highly unlikely that the variations 
is produced by several un-correlated sources.
}
\label{tree}
\end{figure*}
Figure \ref{tree} shows a progressive extension of the Fe K diffuse emission 
of the November 2004, April 2007, March 2008 and April 2009 
observations. 
The Fe K emission seems to evolve with time. The spot on the 
upper right corner of the image (MC 1) does not show clear intensity variations 
(the apparent different morphologies during the 2007, 2008 and 2009 
observations are due to the presence of the EPIC-pn chip edge falling in the 
middle of the source during those pointings). On the other hand significant 
intensity variations appear in the positions traced by the dashed ellipses. 
During the 2004 observation, the bridge 1, 2, 3 and 4 regions are basically 
off and much fainter than MC1 and MC2. In 2007 bridge 1 lights up, becoming
as intense as MC2. Also bridge 2 starts to rise, even if it does not reach 
its maximum yet.
Further increase and evolution occurs in 2008. 
In 2009, it is the bridge 3 that has the higher flux. The time behaviour 
of the phenomenon suggests a connection/evolution in the intensity rise 
of the different regions. Nevertheless, we note that the emitting 
regions are causally disconnected. In fact, the delay time between 
the regions is of the order of 2-4 years, while they are separated by at 
least 15 light years on the plane of the sky, considering their projected 
distance and assuming that they are located at the distance of the GC. 

\section{Discussion}

The deep \xmm\ monitoring of the 15 arcmin radius area around Sgr A* shows 
strong, diffuse and complex pattern of Fe K$\alpha$ emission from neutral 
or low-ionization ions. The Fe K radiation 
is asymmetrically distributed, being brighter along the Galactic Plane and at positive 
galactic longitudes. In particular, a strong correlation between Fe K emitting regions 
and MC is present, as already observed with other telescopes (Sunyaev et al. 1993;
Murakami et al. 2001; Yusef-Zadeh et al. 2007; Koyama et al. 2009). 

The analysis of the \xmm\ data confirms the presence of diffuse emission in this 
region and a correlation with MC. 
An attempt to separate the contribution from the different MC regions has been
done using the CS maps provided by Tsuboi et al. (1999). 
We thus selected the regions shown in Fig. 2 and performed a detailed 
spectral--temporal analysis.
In particular we analysed the X-ray emission in the direction of G0.11-0.11, 
another region that we named "the bridge", plus two other MC regions (MC1 
and MC2). 
We also analysed the spectra of two massive MC close to the Galactic Centre, 
the 20 and 50 km \s\ MC (Coil et al. 2000). 

The spectra of the Fe K$\alpha$ bright MC suggest a reflection origin for this 
emission. The time-resolved spectral variability shows a complex pattern of 
variations, with some MC increasing their Fe K emission, some decreasing, 
and some having a stable Fe K emission. 

\subsection{Superluminal echo}

Figure \ref{tree} shows the evolution of the illumination of the bridge 
region. In 2-5 years, we observe the progressive illumination of a region 
with projected angular size of about 1.92 arc-minutes. The studies of the 
CMZ through the different tracers (i.e., CS and ammonia) show that the 
bridge physical parameters are typical of the MC of the CMZ (Armstrong 
et al. 1985; Morris et al. 1983; Gusten et al. 1981; Tsuboi et al. 1999; 
Amo-Baladron et al. 2009). Indeed the high line-of-sight velocity 
(V$_{\rm LSR}\sim50$ km \s), the high internal temperature and column 
density are consistent with the values observed in the CMZ. 
Moreover, Becklin et al. (1974) and Glass et al. (1987), studying 
the infra-red images of the Galactic Centre, argue that the bridge 
cannot be located closer to us than the Galactic Centre, otherwise it 
would absorb the light coming from the numerous red stars located 
at the Galactic Centre. 
Unlike the 20 and 50 km \s\ clouds, the bridge does not induce
significant extinction in the stellar IR emission, suggesting a position behind  
Sgr A*.

Therefore, the projected angular size of 1.92 arc-minutes 
corresponds to a physical extent of at least 15 light years. 
The variation in the bridge implies, thus, a
superluminal velocity of at least 4 times the velocity of the light.
Fig. \ref{lc_arm14} shows that the light curves of 
the different bridge regions show variations with similar factors,   
intensities and time-scales, but each with a time delay. 
The observation of such 
variations cannot be due to the propagation of an event occurring 
inside the MC because it would have to travel faster than the light 
(if the bridge is sitting on the plane of the sky). 
For the same reason it is also excluded that the observed variations 
are due to the propagation of low energy cosmic rays, produced 
within the molecular cloud.
Moreover, both an internal and a nearby source can be excluded.
In fact, whatever the bridge inclination is, the Fe K intensity variations 
must be decreasing with the square of the distance from the primary 
source. If the source is located either within the MC or nearby, this 
effect must produce, contrary to the observations, a high ratio between 
the intensity variations in the different regions. 
This, thus, indicates that the primary source is more distant than several 
times the bridge length. 
An internal or nearby source can be excluded also because of the lack 
of detection of transient sources, close enough to the bridge, within 
the period of the \xmm\ monitoring. Either null or small delays are expected
if the source is located inside, close or behind, the bridge. 

It is also very unlikely that several different un-correlated events 
happening within the MC (e.g., switching on of weak transients)
can give rise to such a coherent pattern of time and spatial variations
in causally disconnected regions.

Finally, it is interesting to note that in both 2004 observations all the 
bridge regions show significant Fe K emission, even before the fast 
variations occur. This indicates that the bridge was an Fe K emitter 
even before that the super-luminal variation occurred. A single 
burst from a primary source and a peculiar morphology of the reflecting 
material of the bridge might explain this behaviour, nevertheless,
most probably the primary source was active also before that the 
super-luminal  variation occurred.
This suggests that the source underwent a strong and fast short-time 
modulation, but within a longer lasting process. 

In summary, the observation of the superluminal echo, the similar intensity 
Fe K variations in the different regions and the lack of detection of point 
source associated with it, imply that the origin of the variation has to be an 
external source, distant from the bridge. From the derived column density 
(n$_H=9\times10^{22}$ cm$^{-2}$) and assuming a source distance of 18 pc,
(4 times the bridge dimension) we derive, to produce a Fe K line intensity of 
$1.1\times10^{-5}$ ph cm$^{-2}$ \s, a primary X-ray luminosity higher than  
$1.3\times10^{38}$ erg \s, value close to the Eddington luminosity for a Solar 
mass black hole.

Koyama et al. (1996), analysing the X-ray emission from Sgr B2, suggested
that the origin of the Fe K emission in that MC is due to a past flare of Sgr A*.
Sgr B2 is a MC with a column density of $8\times10^{23}$ cm$^{-2}$ 
(see Tab. \ref{MC_par}) and a radius of about 7 pc (Revnivtsev et al. 2004).
Its projected distance is about 100 pc East of Sgr A* and a recent parallax 
measurement places it about 130 pc in front of the plane of Sgr A*
(Reid et al. 2009). Figure \ref{face} shows a sketch of the Galactic plane 
as seen from an observer above the Galaxy. The East of Sgr A* is 
presented with negative values and the Earth is located in the bottom 
of the Figure at negative y- values. Sgr B2 is represented by the blue 
circle. Terrier et al. (2010) measured, in Sgr B2, a Fe K line intensity of 
$1.7\times10^{-4}$ ph cm$^{-2}$ \s. 
Assuming that the line is the echo 
of a past activity of Sgr A*, this implies a Sgr A* flare luminosity 
of L$\sim1.4\times10^{39}$ erg \s\ terminating about 100 years 
ago\footnote{We want to clarify that along all the paper we infer the 
history of the past activity of X-ray sources in the Galactic Centre region 
without considering the lag due to the distance between the Earth and 
Sgr A*. This means that the Sgr A* flare would finish 100 years ago as 
seen from the Earth.}.

The direct implication of this scenario is, however, that the past activity 
of Sgr A* should light up other MC. Are the emission and variations 
of the MC in the central 15 arcmin from Sgr A* consistent with such a
scenario?

\subsection{A common origin for the MC emission and variations:
the glorious past of Sgr A* revealed by MC emission?}
\label{common}

To evaluate this global scenario we have to take into account the 
properties of the different molecular clouds. Table \ref{MC_par} shows
the physical parameters of the MC under investigation.
For most of the MC, the column densities have been measured 
through the integration of the CS emission.

As shown by the work of Amo-Baladron et al. 2009, the n$_H$ derived 
with this method can be over-estimated of factors of the order of a few. 
In fact, this method is integrating the CS emission over the entire line-of-sight 
irrespective of the X-ray morphology. This technique might sum the 
contribution due to clouds having a similar line-of-sight velocity, but not 
associated with the cloud under investigation. Other sources of uncertainty 
are the CS/H abundances ratios and the CS excitation temperature. 
A more detailed study of the MC masses and distribution is beyond the 
scope of the present study, and it will be addressed in future works.

\subsubsection{G0.11-0.11}
\begin{figure}
\vspace{-0.4cm}
\hspace{-0.8cm}
\includegraphics[width=0.55\textwidth,height=0.55\textwidth,angle=0]{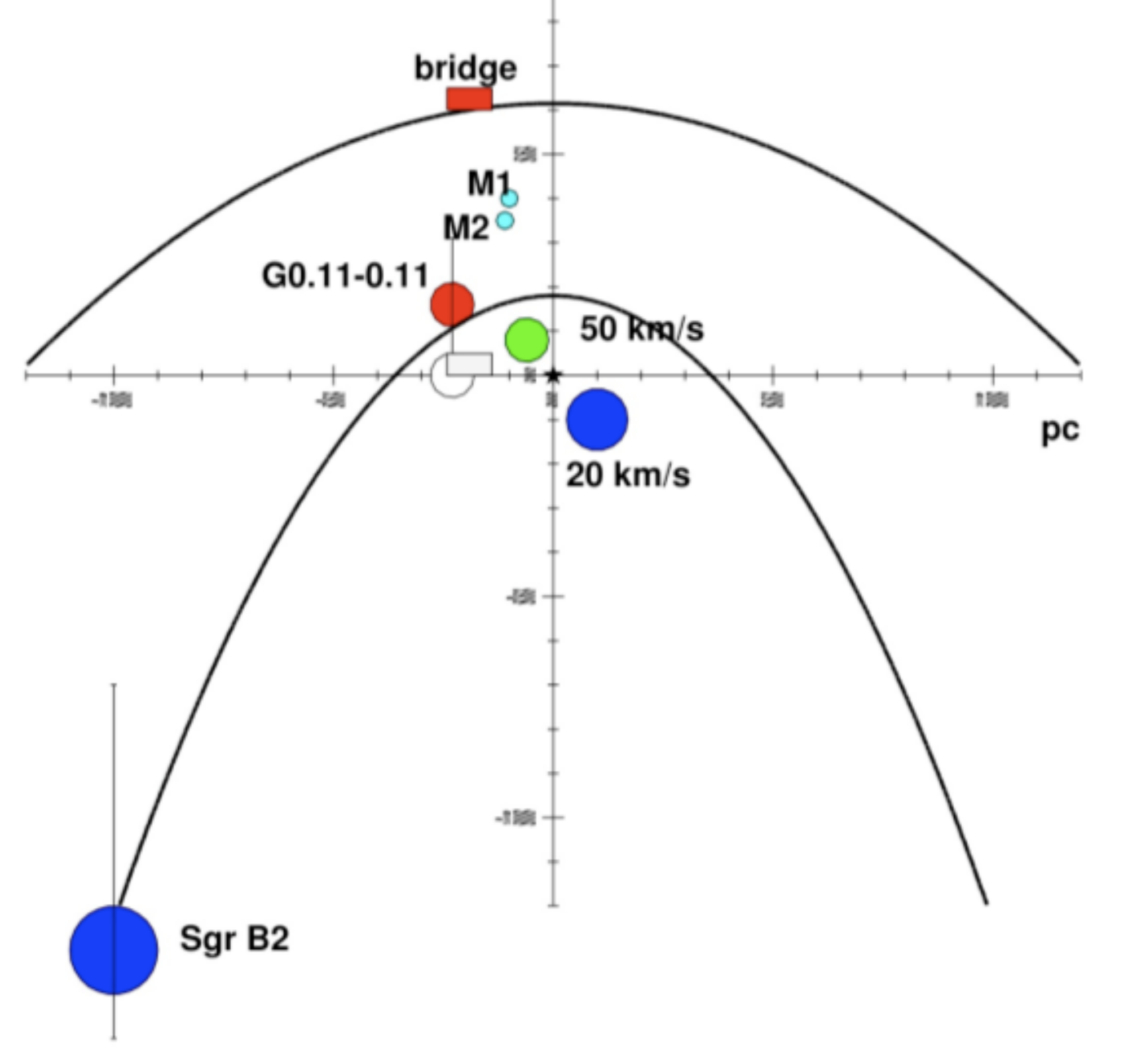}
\caption{A sketch of the molecular face--on view of the Galactic Plane as seen from 
the direction of the north Galaxy pole.
Sgr A* is at the vertex indicated by a black star. Galactic East is toward negative abscissa 
and the direction toward the Earth is bottom, at negative ordinate. Sgr B2 is 
shown with a blue circle about 130 pc in front of Sgr A*. MC1 and MC2 are shown 
by the two light blue circles at 35 and 40 pc, respectively. The 20 and 50 km \s\ 
clouds are presented with blue and green circles. The bridge and G0.11-0.11 
are shown with circles and rectangles, respectively. They are white when 
presented at the projected distance and red when their actual position is estimated.
The parabola hitting Sgr B2 and G0.11-0.11 represent a light front emitted by 
Sgr A* 100 years ago. The further a light front 400 emitted years ago.
}
\label{face}
\end{figure}

The molecular cloud G0.11-0.11 (Tsuboi et al. 1997; Handa et al. 2006) 
is located between the Galactic Center Arc and Sgr A*, at Galactic 
co-ordinates l=0.108$^{\circ}$ b=-0.108$^{\circ}$ (see Fig. \ref{FeK_EW} and
\ref{face}). 
The mass of this MC is highly debated in literature. 
The column density has been estimated through a detailed map in 
H$^{13}$CO$^{+}$ and in thermal SiO lines and resulted to be: 
N(H$_2$)=6-7$\times$10$^{23}$ cm$^{-2}$ (Handa et al. 2006), assuming 
that the H$^{13}$CO$^+$(1-0) emission is thermalized, nevertheless 
Amo-Baladron et al. (2009) showed it to be subthermally excited;
their study of the CS maps indicates a lower column density of 
$5\times10^{22}$ cm$^{-2}$. However, recently a detailed study of 
this MC indicates an even lower value of the column density, with 
N$_H=2^{+1}_{-1.3}\times10^{22}$ cm$^{-2}$ (Amo-Baladron et al. 2009);
we assume this value for all the following calculations.
G0.11-0.11 has a radius of 3.7 pc and a projected distance from Sgr A* of 25 pc.
The strongest feature of the X-ray spectrum is the Fe K line at 6.4 keV. 
It has an intensity of 0.9$\times$10$^{4}$ ph cm$^{-2}$ s$^{-1}$. Assuming 
that the molecular cloud is located at its minimal distance from Sgr A* 
(see the white circle in Fig. \ref{face}) and that the line is due to irradiation 
from it, the flare luminosity has to be L$\geq$10$^{39}$ erg \s\ occurring 
more than 75 years ago (Sunyaev \& Churazov 1998).

These values are surprisingly similar to the Sgr B2 ones. 
Moreover the light curve of both MC is in a decay phase with 
about the same factors of variations (see Inui et al. 2009; Terrier et al. 2010
and Fig. \ref{lc_G011}).
Assuming that G0.11-0.11 is produced by the same flare illuminating 
Sgr B2, it is possible to estimate the distance between Sgr A* and the cloud, 
and thus to measure the distance between G0.11-0.11 and the plane where 
Sgr A* is sitting. Thus, assuming a flare luminosity of 1.4$\times$10$^{39}$ 
ergs s$^{-1}$, this implies that G0.11-0.11 is about 17 pc either in front or 
behind the plane of Sgr A* (red circle in Fig. \ref{face}).

To probe the hypothesis that G0.11-0.11 is responding to the same Sgr A* flare 
illuminating Sgr B2, we can apply another independent constraint. In fact, not only 
their column density, size and Fe K intensities have to be linked, but also their 
distance with respect to Sgr A* has to be tied, in order to respond at the same 
time to the same flare. In other words, the position of G0.11-0.11 has to sit on 
the same illumination front hitting Sgr B2. Figure \ref{face} shows that this is 
exactly the case, if we set G0.11-0.11 behind Sgr A*. 

This is a strong constraint on the single flare hypothesis. It does not 
rule out other possibilities for the Fe K emission, but in those cases 
the observed ratios between line intensities, column densities and 
distances must be such by coincidence. We also foresee a continuum 
correlation between the Sgr B2 and G0.11-0.11 light curves, at least 
if Sgr A* is the dominating source of illumination.

\subsubsection{Other constraints on the recent-past activity 
of Sgr A*: The 20 and 50 km \s\ MC}

Within 15 pc of Sgr A* there are two massive MC, the 20 and 50 km 
\s\ clouds. Thanks to the interactions between these MC and the material around 
them, it is possible to estimate the geometry of the region (Coil et al. 2000).
In particular the 50 km \s\ MC has a projected distance of about 5 pc from 
Sgr A* and seems to be about 5-10 pc behind it (Coil et al. 2000). 
Thus if a strong flare occurred in the recent past, this MC should still be 
reflecting its radiation. 
In Fig. \ref{face} the 50 and 20 km \s\ MC are represented with the 
green and blue circles, respectively. The CS maps indicate, for the 50 
km \s\ MC, a column density of about n$_H=9\times10^{22}$ cm$^{-2}$.
Thus, the upper limit on the neutral Fe K line places a limit on the mean 
luminosity produced by Sgr A* in the past 60-90 years (depending on the 
exact location), that is L$\leq8\times10^{35}$ erg \s. 
The study of Sgr B2, G0.11-0.11, the 20 and 50 km \s\ MC suggests, thus,
that Sgr A* was in an active state about 100 years ago (L$\sim10^{39}$ erg 
\s). Then, Sgr A* 
entered in a quiet period of low activity (with mean luminosity
lower than several 10$^{35}$ erg \s) until now. In this interpretation, we expect 
that the X-ray luminosity of Sgr B2 and G0.11-0.11, induced by Sgr A*, 
should be significantly lower in the next few decades.

\subsubsection{The Bridge}

As well as Sgr B2 and G0.11-0.11, the bridge is consistent with being 
illuminated by an external source but, opposite to the other MC, 
its light curve shows a clear increase. Thus these MC cannot be 
seeing the same phase of the flare illuminating Sgr B2.

The observation of the apparent super-luminal motion indicates 
an external illuminating source as the source of irradiation. 
As derived in paragraph 7.1, the source must have a luminosity 
higher than $1.3\times10^{38}$ erg \s.
This value is close to the Eddington luminosity for a stellar mass 
black hole. 
We note that, although accretion onto a neutron star can reach even 
higher luminosity for short periods, it is improbable that it steadily 
produces such radiation for periods of years. 
On the other hand, a black hole X-ray binary can produce 
such a luminosity for several years. However, the classical duty cycle 
of a X-ray binary outburst is a few months to years, thus we would 
expect modulations on month time scales and a decrease of the Fe K 
emission in the next years. Further, deep X-ray observations of the 
field are thus mandatory to clarify this issue.

We also stress that we do not observe any bright enough X-ray
source during the \xmm\ monitoring around the bridge region. 
The brightest transient, active during the \xmm\ monitoring 
and close to the bridge, is XMM J174554.4-285456, which has an 
estimated luminosity of about $2\times10^{34}$ erg \s, not able to 
produce the observed variation. 
This implies that the delay between the arrival time of the primary 
and the reflected emission has to be significant. This excludes that 
the primary source is placed behind the cloud. 
Assuming that the maximal luminosity produced by a binary is of the 
order of 10$^{39}$ erg \s, we can place an upper limit to its distance 
from the bridge of the order of 50 pc. We thus constrain the position 
of a possible binary responsible for the bridge variation to a distance 
between 18 and 50 pc from the bridge.
Moreover, we note that (see Fig. \ref{tree}) the illumination starts to the 
Galactic west of the bridge and propagates towards the Galactic east, 
diffusing in the northern part of the bridge. This strongly suggests that the 
illuminating source is located east of the bridge, slightly north of it. 
This is the direction toward which Sgr A* is located.

The bridge might, in fact, be reflecting the light produced by 
an active period of Sgr A*. If the bridge is at its minimal distance 
from Sgr A* of about 18 pc, the flare luminosity should be 
$\sim10^{38}$ erg \s. Assuming the same luminosity as seen 
by Sgr B2 and G0.11-0.11 of L$=1.4\times10^{39}$ erg \s, 
we estimate the bridge location at about 60 pc 
behind the plane of Sgr A* (it cannot be in front because in that 
case the 50 km \s\ cloud should see the same flare). 
The location of the bridge about 60 pc behind Sgr A* is in agreement 
with the weak infrared absorption of the Galactic Centre stars 
observed in the direction of the bridge (G\"usten et al. 1980; 
Glass et al. 1987). This position well behind Sgr A* is in agreement 
with an active phase of Sgr A* that lasted about 400 years.

\subsubsection{MC1 and MC2}

The two MC called MC1 and MC2 have intense, but stable Fe K 
emission. This confirms the result by Muno et al. (2007) who find a 
constant Fe K intensity coming from these regions. Nevertheless, the 
superb imaging capabilities of \chandra\ allowed the detection of 
changes in the shape of the emitting regions (Muno et al. 2007).

MC1 corresponds to the MC observed in ammonia and CO 
and called M0.02-0.05 by Armstrong et al. (1985). 
Due to its small line-of-sight velocity, the position of this MC is 
still debated. Nevertheless the strong Fe K emission and the broad
molecular line-widths suggest a location within the CMZ for MC1. 
MC1 has a projected distance 
of 12.5 pc from Sgr A*, a column density of 4$\times$10$^{22}$ 
cm$^{-2}$ (estimated using the CS maps), a radius of about 1.8 pc, 
and a Fe K line intensity of 2$\times$10$^{-5}$ ph cm$^{-2}$ \s.
As for the bridge, a flare from Sgr A* must have had a luminosity 
of about 7$\times$10$^{37}$ ergs s$^{-1}$ to illuminate MC1, if located 
at the projected distance. Again, arbitrarily assuming a Sgr A* flare 
luminosity of $1.4\times10^{39}$ erg \s, implies that MC1 must be 
about 50 pc behind Sgr A* (see the light blue circle in Fig. \ref{face}).

In the position of the MC2 cloud there are not strong CS over-densities.
We observe a CS flux corresponding to a column density of n$_H<2
\times10^{22}$ cm$^{-2}$. The projected distance of MC2 corresponds 
to 14 pc, the radius to 1.8 pc and the line intensity to $1.2\times$10$^{-5}$ 
ph cm$^{-2}$ \s. Fig. \ref{face} shows that if MC2 is illuminated by
the same flare of Sgr A* as Sgr B2, it should be located less than about 35 
pc behind the super-massive black hole.
If this is the case, we predict that this MC should soon enter in the decay 
phase that Sgr B2 and G0.11-0.11 are experiencing now.

\subsection{The glorious (or in-glorious?) past of Sgr A* betrayed by MC}

The X-ray emission from Sgr B2 and the 50 km \s\ MC indicate that 
the Sgr A* luminosity was lower than $8\times10^{35}$ erg \s\ until
60-90 years ago, while it was brighter about 100 years ago with at least 
10 years of nearly constant emission (with a luminosity of about 
$1.4\times10^{39}$ erg \s). 

The uncertainty on the position of the other MC do not allow to derive 
strong constraints on the Sgr A* luminosity. Nevertheless, we can pose 
an upper limit to the distance between Sgr A* and these MC of 200-300
pc (that correspond to the size of the CMZ).
This immediately can be translated into an upper limit to the luminosity 
of Sgr A*, in the last few thousand of years, of L$\sim10^{41}$ erg \s. 
This is in agreement with previous results obtained studying the 
X-ray emission from the MC in the Galactic disc (Cramphorn \& 
Sunyaev 2002). Thus, Sgr A*, even if brighter, did not reach accretion 
rates higher than $\eta\sim10^{-5}$ the Eddington rate in the recent past.

In Figure \ref{face} we estimated the MC positions arbitrarily assuming 
a Sgr A* luminosity of $1.4\times10^{39}$ erg \s\ for many years. 
In reality Sgr A* might have had, in the past, many different flares with 
different peak luminosity, instead of one single period of activity. 
Only when the distance of the MC can be measured in an independent
way, it will be possible to safely estimate the luminosity of the Sgr A* 
activity and discriminate between these two scenarios. 
Nevertheless, a clue on this issue can be obtained through the comparison 
with similar black hole accretion systems. 
The Sgr A* accretion rate implied by the Sgr B2 emission corresponds 
to the ones of the Galactic Black Holes (GBH) in the transition between 
quiescence and low hard state. 
Assuming that the Power Spectral Density of Sgr A*, at that accretion 
rate, is similar to the one of GBH in the low hard state, with just the 
time-scale of the variations scaled with the different BH mass, then 
for Sgr A* a low level of variations would be expected on a time-scale 
longer than years (GBH show little variability at frequencies lower than 
0.01 Hz, Remillard \& McClintock 2006). For this reason a long period 
of activity might be preferred to many strong flares. 
In this hypothesis, we would witness the rising phase (through the 
bridge) and the decay phase (through Sgr B2 and G0.11-0.11). 
In particular, the active phase of Sgr A* should have lasted 
about 400 years. This implies that a significant number of 
MC should still be reflecting the Sgr A* radiation. In particular, 
assuming a uniform distribution of the MC within the CMZ and that 
the CMZ can be described by a disc with radius of 200 pc, 
it is possible to estimate (by computing the surface included 
between the two parabolas describing the beginning and the 
end of the flare) that nowadays we should observe about 30 \% 
of the MC in the CMZ still reflecting Sgr A* emission. 
Interestingly this value is 
of the same order of magnitude of the ratio between X-ray bright 
and weak MC that can be roughly estimated from the 
comparison between the Fe K and molecular emission surveys 
of the CMZ (Yusef-Zadeh et al. 2007; Koyama et al. 1996). 
However more detailed work is needed for a better estimate of
this ratio and the comparison between different outburst models.

\subsection{Alternatives to Sgr A*}

The observation of a superluminal echo rules out either an internal source 
or cosmic rays as being responsible of the Fe K emission for the bridge.
Moreover, the similarity between the Fe K intensity and MC physical 
properties of G0.11-0.11 and Sgr B2 suggests that an external X-ray 
source should have a luminosity of at least 10$^{39}$ erg \s\ for many 
years and to be located close (a few parsecs) to the Galactic Centre.
If we assume a unique event to explain the Fe K luminosity of all the MC,
then the only possible alternative, to a flare from Sgr A*, is the interaction 
between Sgr A East and the 50 km \s\ molecular cloud. 
Fryer et al. (2006) calculated, in fact, that the impact between the shock 
produced by the supernova and the massive molecular cloud might 
have produced a luminosity up to about 10$^{39}$ erg \s\ a few 
hundred years ago, with a considerably long period of activity.
 
Nevertheless the fast variability observed in the bridge region seems 
at odds with such a scenario. In fact between 2008 and 2009, the 
bridge 3 region increased its Fe K emission by a factor of 2.
This implies that the emitting region must be smaller than about 0.3 pc.
Therefore, dense matter clumps with light-years-scales must dominate 
the emission over an extended emission produced by the interaction 
between the 50 km \s\ MC and the Sgr A East shock. 
Moreover, the model reported in Fryer at al. (2006) predicts 
a slow decay of the emission produced by the Sgr A East shock.
This might be difficult to reconcile with the factor of 2 decrease 
observed in G0.11-0.11 in 8 years and the four order of magnitude 
decrease of the emission in the 50 km \s\ MC in only about 100 years.
Thus, if only one source is illuminating all the MC, then Sgr A* is the 
most probable candidate. 

Nevertheless, we know that the GC region often experiences outbursts from 
transients. Some of these have high X-ray luminosity for a considerable amount 
of time and they may be placed in the correct position to highlight a nearby MC. 
At present, in fact, it cannot be excluded that the variation seen in the bridge 
region is due to an X-ray transient. However, a bright enough X-ray transient 
has never been observed close to the bridge, to G0.11-0.11 or to Sgr B2. 
In this interpretation, the transients should have been active 
before that the X-ray monitoring started. This implies that the illuminating 
sources must always be located in front of the MC, otherwise no big delay would 
be observed. Although possible, we think that this is unlikely. 

As pointed out by Yusef-Zadeh et al. (2002; 2007) the X-ray bright MC in the 
GC are very well correlated with the non-thermal radio emission. A distribution 
of low energy cosmic rays can produce low ionisation Fe K lines with significant 
EW (of the order to 200-300 eV). Even if at present the Fe K emission from 
Sgr B2 and G0.11-0.11 is dominated by the radiation field produced by Sgr A*,
it is important to continue to monitor those MC because the prediction is that 
they should switch off in the near future. Then, the Fe K emission induced 
by the low energy electrons, if any are present, should persist, and become 
the dominating emission mechanism.

\section{Conclusions}

The deep \xmm\ 8 years monitoring of the Sgr A* 
region has allowed us to carry out a detailed study of the 
Fe K$\alpha$ emission within the central 15 arcmin 
of the Galactic Centre.
We focussed on the main emission features and reached 
the following conclusions:

\begin{itemize}

\item{} The Fe K$\alpha$ emission is asymmetrically distributed
and is concentrated in features that are correlated to some high density
molecular clouds of the CMZ. These X-ray Fe K bright MC show all the 
features of a reflection spectrum: neutral Fe K lines with EW 
of about 0.7-1 keV, Fe K edges ($\tau\sim0.1-0.3$) and flat 
continua ($\Gamma\sim$0.6-1.7).

\item{} We discovered an apparent super-luminal motion 
corresponding to a MC called "the bridge". The illumination 
front spans at least 15 light years in a time-scale of about 2-4 
years. This phenomenon cannot be due to a source internal 
to the molecular cloud or to propagation of low-energy cosmic rays. 
A series of events is also unlikely because a similar pattern
of variability is observed in causally disconnected regions. 
The most probable cause of this variation is the illumination 
of the molecular material from an external source. 
One possible geometry is that the illuminating fronts are nearly 
parallel to the bridge. To produce an apparent super-luminal 
motion and the observed pattern of variations, the X-ray radiation 
must originate far away from the reflector (more than about 18 pc). 
The source must thus have had a luminosity higher than about 
$1.3\times10^{38}$ erg \s\ for several years. 
A prediction of this picture is that the variations observed in the 
bridge region 1 propagate to region 2 and then 3 (maybe bridge 
region 4).
Assuming that Sgr A* is the primary source and that it had a 
luminosity of about 10$^{39}$ erg \s, it would place the bridge about 
60 pc behind Sgr A*. This would imply either a long (about 400 years) 
or, to take intermittency  to the extreme, it could imply that there have 
been multiple (i.e., two) brief events of activity of Sgr A* in the past. 

\item{} We observe that the Fe K emission from G0.11-0.11 is 
decreasing, as if the MC is responding to the same flare that is 
illuminating Sgr B2, whose emission is also fading (Inui et al. 2009; 
Terrier et al. 2010). 
Considering the sizes and column density of the MC we 
estimate for G0.11-0.11 a position of about 17 pc behind Sgr A*,
from the intensity of the Fe K line only. 
Surprisingly, in that position, G0.11-0.11 satisfies another completely 
independent constraint. In fact, it is consistent with being 
illuminated by the same light front hitting Sgr B2. 
These two MC are, thus, experiencing the decay 
of a period of activity of Sgr A* (activity characterised by a 
luminosity of about $1.4\times10^{39}$ erg \s) that lasted for
at least 10 years (Revnivtsev et al. 2004) and that ended about 
100 years ago (between 70 and 150 years ago).
Other possibilities for the Fe K emission are not excluded, 
but in those scenarios the observed properties would result from 
pure chance.

\item{} The upper limit on the neutral Fe K emission from 
the 20 and 50 km \s\ molecular clouds allow us to further constrain 
the mean level of activity of Sgr A* in the recent past, being lower 
than 8$\times10^{35}$ erg \s\ in the last 60-90 years. 
The Fe K emission from Sgr B2, G0.11-0.11 and the 50 km \s\ 
cloud suggest that the emission (presumably induced by Sgr A*) 
from Sgr B2 and G0.11-0.11 is destined to switch off in the next decades. 
At that point the emission produced by low energy cosmic rays, 
if present, might become dominant.

\item{} We also observe two bright molecular clouds with intense 
but stable Fe K emission. Arbitrarily assuming that they reflect 
a flare of 10$^{39}$ erg \s, they would be placed at about 35 
and 40 pc behind Sgr A*. 

\item{} The emission and variations of the MC in the central part 
of the Galaxy fit the scenario of a past period of activity of Sgr A*
(either continuous or with bursts of emission). 
This period might have started a few hundreds of years ago and lasted 
until about 70-150 years ago. Since then, Sgr A* experienced a long period 
of low activity until now. 
This idea is in agreement with the observation of many Fe-K-bright MC 
as well as many X-ray weak MC and with the (poorly) known 3-d 
distribution of MC within the CMZ. 

\item{} The analysis of the long \xmm\ monitoring of the Galactic Centre 
shows that, in the past, Sgr A* had activity levels of the order of 
$\eta=10^{-5}$ of its Eddington luminosity (as compared to its current 
$\eta=10^{-8}$). 
In those conditions it appeared more similar to the classical low 
luminosity AGN and the Galactic Black Holes in the quiescent state. 
A deeper study of that period is important for the 
comprehension of the accretion mechanism, in view of the unification 
schemes.

\end{itemize}

Continuous deep X-ray monitoring of the region, as well
as an improvement of the molecular maps of the CMZ are mandatory 
to fully understand the origin of the X-ray emission from the MC, 
in particular, to disentangle the contribution to the observed variations 
to be ascribed to the supermassive black hole of the galactic center 
and to X-ray binaries of the region. 

The connections between the X-ray Fe K maps, their variations and 
the measurements of the column densities and positions of the 
MC within the CMZ will allow to precisely reconstruct
the history of the emission of Sgr A* and of the bright X-ray sources
inside the CMZ. The flares from these sources will provide a tool 
to scan the CMZ and reconstruct the small scale distribution of MC. 

\section*{Acknowledgments}

The work reported here is based on observations obtained with
XMM-Newton, an ESA science mission with instruments and contributions
directly funded by ESA Member States and NASA. The authors 
wish to thank the referee Prof. Mark Morris for the numerous comments 
that greatly helped to improve the paper. GP thanks ANR for 
support (ANR-06-JCJC-0047). GP wish to thank M. Tsuboi for kindly providing 
the CS maps, Anne Decouchelle, Vincent Tatischeff, Jesus Martin-Pintado Martin, 
Massimo Cappi, Stefano Bianchi, Matteo Guainazzi, Piergiorgio Casella, 
Kallmann Nagi and Silvia Boscherini for useful discussion.

\end{document}